\documentclass[aip,reprint]{revtex4-1}

\usepackage{booktabs}
\usepackage{filecontents}
\usepackage{url}
\usepackage{amssymb,amsmath}
\usepackage{graphicx}
\usepackage{color,soul}

\begin{document}


\title{GMSCN: Generative Model Selection Using a Scalable and Size-Independent Complex Network Classifier} 



\author{Sadegh Motallebi}
\email[]{motallebi@ce.sharif.edu.}
\affiliation{Department of Computer Engineering, Sharif University of Technology, Tehran, Iran}

\author{Sadegh Aliakbary}
\email[]{aliakbary@ce.sharif.edu.}
\affiliation{Department of Computer Engineering, Sharif University of Technology, Tehran, Iran}

\author{Jafar Habibi}
\email[]{jhabibi@sharif.edu.}
\affiliation{Department of Computer Engineering, Sharif University of Technology, Tehran, Iran}

\date{\today}

\begin{abstract}
Real networks exhibit nontrivial topological features such as heavy-tailed degree distribution, high clustering, and small-worldness. Researchers have developed several generative models for synthesizing artificial networks that are structurally similar to real networks. An important research problem is to identify the generative model that best fits to a target network. In this paper, we investigate this problem and our goal is to select the model that is able to generate graphs similar to a given network instance. By the means of generating synthetic networks with seven outstanding generative models, we have utilized machine learning methods to develop a decision tree for model selection. Our proposed method, which is named ``Generative Model Selection for Complex Networks" (GMSCN), outperforms existing methods with respect to accuracy, scalability and size-independence.
\end{abstract}



\maketitle 

\begin{quotation}
A realistic network generative model can generate artificial graphs similar to real networks. But there is no “best universal generative model” for all situations. In any application, among different existing generative models, we should choose the best model for that specific application. So, “generative model selection” is a prerequisite for creating artificial realistic networks. In this paper, we consider the problem of model selection and propose a method for finding the model that best fits a given network. The selected generative model helps us infer the growth mechanisms of the given network. It can also generate artificial networks similar to the given network for tasks like simulation, prediction, extrapolation, and hypothesis testing. We propose utilizing a combination of different local and global network features for learning a model selection decision tree. We also devise a new network feature based on the quantification of the degree distribution. We show that Our proposed method is robust, scalable, independent from the size of the given network, and more accurate than the baseline method.
\end{quotation}

\section{\label{Introduction}Introduction}

Complex networks appear in different categories such as social networks, citation networks, collaboration networks, and communication networks \cite{Newman, StatisticalMechanics,SurveyOfMeasurements, GraphConcepts}. In recent years, complex networks are frequently studied and many evidences indicate that they show some non-trivial structural properties \cite{Newman, SurveyOfMeasurements, StructDynam, DynamicalSys, WaterDistribution, MusicRecom}. For example, power law degree distribution, high clustering and small path lengths are some properties that distinguish complex networks from completely random graphs. \\

An active field of research is dedicated to the development of algorithms for generating complex networks. These algorithms, called ``generative models", try to generate synthetic graphs that adhere the structural properties of complex networks \cite{SurveyOfMeasurements,StatisticalMechanics}. 
Realistic generative models have many applications and benefits. Once a generative model is fitted to a given real network, we can replace the real network with artificial networks in tasks such as simulation, extrapolation (by generating similar graphs with larger sizes), sampling (reverse of extrapolations), capturing the network structure and networks comparison \cite{Kronecker, RTG}.
\\
Despite the advances in the field, there is no universal generative model suitable for all network types and features. The prerequisite of network generation is the stage of generative model selection. In fact, when we generate synthetic networks, we hope to reach graphs that are structurally similar to a target network. In the model selection stage, the properties of a given network (called \textit{target network}) are analyzed and the best model suitable for generating similar networks is selected. A model selection method tries to answer this question: ``Among candidate generative models, which one is the most suitable one for generating complex network instances similar to the given network?" In this paper, we investigate this problem and by the means of machine learning algorithms, we propose a new model selection method based on network structural properties. The proposed method is named ``Generative Model Selection for Complex Networks" (GMSCN).
The need for model selection is frequently indicated in literature \cite{ModelSelection, NetSamplingClassification, Drosophila}. More specifically some works \cite{ModelSelection, Drosophila, Superfamilies} are based on counting subgraphs of small sizes (called \textit{graphlets} or \textit{motifs} \cite{Motifs, Superfamilies, GraphConcepts, SubgraphCounting, NetMotifDiscovery, EngineeringView, EffectOfNetTopology, ModelSelection, Drosophila}), and some others concentrate on structural features of complex networks \cite{NetSamplingClassification}, and some are based on manually selecting a model through watching a small set of network features \cite{RichClub}. We will show that by using an appropriate combination of local and global network features, we can develop a more accurate model selection method. In our proposed method (GMSCN), we consider seven prominent generative models by which we have generated datasets of network instances. The datasets are used as training data for learning a decision tree for model selection. Our method also consists of a special technique for quantification of degree distribution. In comparison to existing methods \cite{ModelSelection, Drosophila, NetSamplingClassification}, we have considered wider, newer and more significant generative models. Due to a better selection of network features, GMSCN is also more efficient and more scalable than similar methods \cite{ModelSelection, Drosophila}.
\\
The rest of this paper is organized as follows. Section \ref{RelatedWorks} reviews the related work. Section \ref{ProposedMethod} presents GMSCN. Section \ref{Evaluation} is dedicated to evaluation of GMSCN. Section \ref{CaseStudy} describes a case study on some real network samples. The results and evaluations of this paper are discussed in Section \ref{Discussion}. Finally, Section \ref{Conclusion} concludes the paper.

\section{\label{RelatedWorks}Related Work}

\subsection{Network Generation Models}
In this subsection, we briefly introduce the leading methods of network generation:

\begin{itemize}
\item\textit{Kronecker Graphs Model (KG)}  \cite{Kronecker}. This model generates realistic synthetic networks by applying a matrix operation (the kronecker product) on a small initiator matrix. This model is mathematically tractable and supports many network features such as small path lengths, heavy tail degree distribution, heavy tails for eigenvalues and eigenvectors, densification and shrinking diameters over time.

\item \textit{Forest Fire Model (FF)} \cite{GraphsOverTime}.  In this model, edges are added in a process similar to a fire-spreading process. This model is inspired by Copying model \cite{WebAsGraph} and Community Guided Attachment \cite{GraphsOverTime} but supports the shrinking diameter property.

\item\textit{Random Typing Generator Model (RTG)}  \cite{RTG}. RTG uses a process of ``random typing" for generating node identifiers. This model mimics real world graphs and conforms to eleven important patterns (such as power law degree distribution, densification power law and small and shrinking diameter) observed in real networks \cite{RTG}. 

\item \textit{Preferential Attachment Model (PA)} \cite{EmergenceOfScaling}. The classical preferential attachment model generates scale-free networks with power law degree distribution. In this model, the nodes are added to the network incrementally and the probability of the attachments depends on the degree of existing nodes.

\item\textit{Small World Model (SW)} \cite{SmallWorld}. This is another classical network generation model that synthesizes networks with small path lengths and high clustering. It starts with a regular lattice and then rewires some edges of the network randomly.

\item \textit{Erd{\"o}s-R\'enyi  Model (ER)} \cite{ER}. This model generates a completely random graph. The number of nodes and edges are configurable.

\item \textit{Random Power Law Model (RP)} \cite{RP}. The RP model generates synthetic networks by following a variation of ER model that supports the power law degree distribution property.
\end{itemize}

Other generative models are also available (we have not utilized them but they are used in related model selection methods), such as Copying Model (CM) \cite{WebAsGraph}, Random Geometric Model (GEO) \cite{RandomGeoGraphs}, Spatial Preferential Attachment (SPA) \cite{SpatialWebGraph}, Random Growing (RDG) \cite{RandomlyGrown}, Duplication-Mutation-Complementation (DMC) \cite{ProteomeEvolution}, Duplication-Mutation using Random mutations (DMR) \cite{RandomlyGrown}, Aging Vertex (AGV) \cite{HighlyClustered}, Ring Lattice (RL) \cite{RandomGraphs}, Core-periphery (CP) \cite{CorePeriphery}, and Cellular model (CL) \cite{CellularNets}.

\subsection{Model Selection Methods}
The aim of this paper and the model selection methods is to find the best generative model that fits a given network instance. Some model selection methods are based on graphlet counting \cite{ModelSelection, Drosophila, Superfamilies}. Graphlets are subgraphs of bounded sizes (e.g., all possible subgraphs with three or four nodes) and the frequency of graphlets in a network is considered as a way of capturing the network structure \cite{ModelSelection}. In some works, directed graphs and graphlets are considered \cite{Superfamilies, Motifs} and some others consider the network as simple (undirected) graphs \cite{ModelSelection, Superfamilies}.\\

Janssen et al. \cite{ModelSelection} have tested both graphlet features and structural features (degree distribution, assortativity and average path length) in the model selection problem. They conclude that counting graphlets of three and four nodes is sufficient for capturing the structure of the network, i.e., appending structural features to the feature vector of graphlet counts does not improve the accuracy of the model selector. In this paper, we critique this claim and show that using a better set of local (such as transitivity) and global (such as effective diameter \cite{GraphEvolution, GraphsOverTime}) network structural features, along with an appropriate degree distribution quantification algorithm, actually improves the accuracy of the model selection. In fact, graphlet counts are limited local features and are not able to reflect the structural properties of a network instance. Janssen et al \cite{ModelSelection} implemented six generative models and generated a dataset of synthetic networks as the training data for decision tree learning \cite{MulticlassADT}. In this method, candidate generative models are: PA \cite{EmergenceOfScaling}, CM \cite{WebAsGraph}, GEO \cite{RandomGeoGraphs} (GEO2D and GEO3D) and SPA \cite{SpatialWebGraph} (SPA2D and SPA3D).
\\
A similar method is proposed by Middendorf et al. \cite{Drosophila}. In this method, the feature vectors are the counts of graphlets with small sizes. Seven different generative models are considered by which network instances are generated as the training data. Candidate generative models are: ER \cite{ER}, PA \cite{EmergenceOfScaling} , SW \cite{SmallWorld}, RDG \cite{RandomlyGrown} , DMC \cite{ProteomeEvolution}, DMR \cite{RandomlyGrown} and AGV \cite{HighlyClustered}. The authors have used a generalized decision tree called alternating decision tree (ADT) as the learning algorithm.
\\
Airoldi et al. \cite{NetSamplingClassification} propose to form feature vectors according to structural network properties. They have considered some classical generative models and generated a dataset by which a na{\"i}ïve Bayes classifier is learned. Candidate generative models are: PA \cite{EmergenceOfScaling}, ER \cite{ER}, RL \cite{RandomGraphs}, CP \cite{CorePeriphery} and CL \cite{CellularNets}. This method is dependent on the size and average connectivity of the target network and this dependency is one of its limitations.\\

Patro et al. \cite{MissingModels} propose a framework for implementing network generation models. The user of this framework can specify the important network features and the weight of each feature. In other words, we consider each generative model as a class of networks. This model, more than to be a specific method, is a relatively open framework and the user should determine different parameters of the framework according to the target application.\\

\section{\label{ProposedMethod}The Proposed Method}
GMSCN is based on learning a classifier for model selection. The goal of a classifier is to accurately predict the target class for a given network instance and in our method, generative models play the role of network classes. In GMSCN, the classifier suggests the best model that generates networks similar to a given network. The inputs of the classifier are the structural properties of the target network and the output is the selected model among the candidate network generation models.
\subsection{\label{subsec:Methodologys}Methodology}
\begin{figure}
\includegraphics{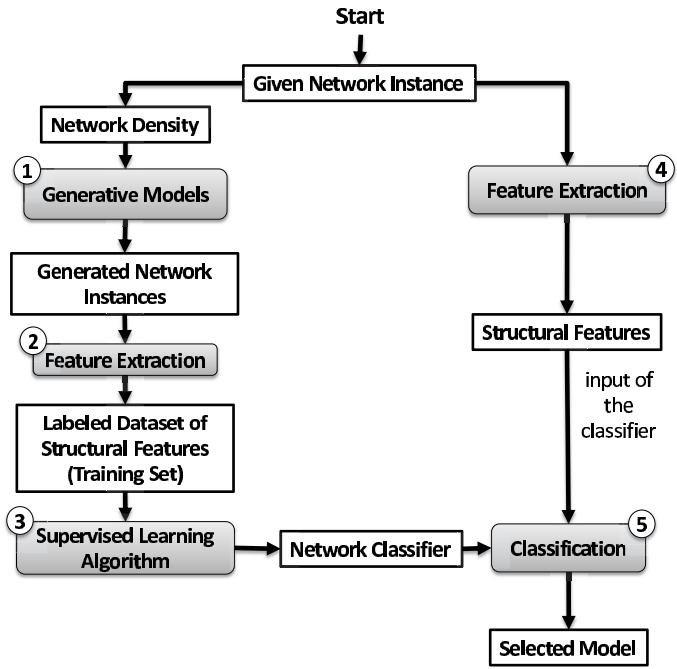}
\caption{ \label{fig:Methodology} The methodology of learning a network classifier}
\end{figure}

Fig. \ref{fig:Methodology} shows the high-level methodology of GMSCN. The methodology is configurable by several parameters and decision points, such as the set of considered network features, the chosen supervised learning algorithm and the candidate generative models. The steps of constructing the network classifier, as illustrated in Fig. \ref{fig:Methodology}, are described in the following:

\begin{enumerate}
\item	Many artificial network instances are synthesized using the candidate network generative models. These network instances will form the dataset (training and test data) for learning a network classifier. In this step, the parameters of the generative models are tuned in order to synthesize networks with densities similar to the density of the given target network. 
\item	After generating the network instances, the structural features (e.g., the degree distribution and the clustering coefficient) of each network instance are extracted. The result is a dataset of labeled structural features in which each record consists of topological features of a synthesized network along with the label of its generative model. 
\item	The labeled dataset forms the training and test data for the supervised learning algorithm. The learning algorithm will return a network classifier which is able to predict the class (the best generative model) of the given network instance.
\item	The structural features of the target network are also extracted. The same ``Feature Extraction" block which is used in the second step is applied here. The structural features of the target network are used as input for the learned classifier.
\item	The learned network classifier is a customized ``model selector" for finding the model that fits the target network. It gets the structural features of the target network as input and returns the most compatible generative model.
\end{enumerate}

In this methodology, the density of the target network is considered as an important property of the target network. Network density is defined as the ratio of the existing edges to potential edges and is regarded as an indicator of the sparseness of the graph. In the proposed methodology, generative models are configured to synthesize networks with densities similar to the density of the target network. This decision is due to the fact that it is hard to compare networks of completely different densities for predicting their growth mechanism and generation process. On the other hand, even with similar network densities, various generative models create different network structures. So, we try to keep the density of the generated networks similar to the density of the target network. In this manner, the network classifier can learn the difference among the structure of various generative models with similar network densities. \\

It is also worth noting that it is not possible to generate networks with exactly equal densities with some of the existing generative models. This is because some generative models (such as Kronecker graphs and RTG) are not configurable for finely tuning the exact density of synthesized networks. So, we generate the networks of training data with similar, and not exactly equal, densities to the density of the given network.\\

Our proposed methodology, unlike existing methods \cite{ModelSelection,NetSamplingClassification,Drosophila}, is not dependent on the size (number of nodes) of the target network. Size-independence is an important feature of our method. It enables the classifier to learn from a dataset of generated networks with sizes different -perhaps smaller- from the size of the target network, but with a similar density. This facility decreases the time of network generation and feature extraction considerably. We will demonstrate the size-independence property of the GMSCN in the evaluation section.\\

GMSCN is actually a realization of the described methodology. In the following subsections, we further illustrate the details of GMSCN by specifying the open parameters and decision points of the methodology.

\subsection{\label{subsec:NetworkFeatures}Network Features}
The process of model selection, as described in Fig. \ref{fig:Methodology}, utilizes structural network features in the second and fourth steps. There are plenty of different network features, so we clarify the considered features in GMSCN here.\\

To capture the properties of a network, we should analyse a wide and diverse feature set of network connectivity patterns. We propose the utilization of a combination of local and global network structural features. The utilization of a limited set of local features (graphlet counts) in similar methods \cite{ModelSelection,Drosophila} has resulted in a lower precision for the model selector. As explained later, we have utilized ten network features from four feature categories. While trying to find the best and minimal set of network features, we considered features that are not only effective on the classification accuracy, but also efficiently computable and size-independent. One may consider a longer list of network features, even from different feature categories (e.g. eigenvalues). In such an approach, automatic methods for feature selection such as the methodology explained in Ref. \cite{Zanin} may be helpful. But supporting specified diverse criteria (effectiveness, efficiency and size-independence) for selected features is quite difficult in such an automatic methodologies.
\\
The utilized features and measurements in GMSCN are:

\begin{itemize}

\item	\textit{Transitivity of relationships}. In this category of network features, we consider two measurements of ``average clustering coefficient" \cite{SmallWorld, Newman} and ``transitivity" \cite{TransitivityProp}.

\item \textit{Degree correlation}. The measure of assortativity \cite{Newman} is selected from this category of network features.

\item \textit{Path lengths}. There are different global features about the path lengths in a network, such as diameter \cite{DiameterProp}, effective diameter \cite{GraphEvolution, GraphsOverTime} and average path length \cite{SurveyOfMeasurements}. We selected the ``effective diameter" measurement since it is more robust \cite{Kronecker} and also because of its less computation cost and sensitivity to small network changes \cite{Sensitivity}. Effective diameter indicates the minimum number of edges in which 90 percent of all connected pairs can reach each other \cite{Kronecker, GraphEvolution, InternetTopology}. Effective  diameter is well defined for both connected and disconnected networks \cite{GraphEvolution}.

\item	\textit{Degree distribution}. It is a common approach to fit a power law on the degree distribution and extract the power law exponent as a representative quantity for the degree distribution. But a single number (the power law exponent) is too limited for representing the whole degree distribution. On the other hand, some real networks do not conform to the power law degree distribution \cite{Slashdot, GooglePlus, WhatIsTwitter}. We propose an alternative method for quantification of the degree distribution by computing its probability percentiles. The percentiles are calculated from some defined regions of the degree distribution according to its mean and standard deviation. We devise $K$ intervals in the degree distribution and then calculate the probability of degrees of each interval. $K$ is always an even number greater than or equal to four. The size of all intervals, except the first and the last one, is considered equal to $p\sigma$ where $\sigma$ is the standard deviation of the distribution and $p$ is a tunable parameter. In any application, we can configure the values of $K$ and $p$ in a manner that the percentile values become more distinctive. In our experiments we let $K=6$ and $p=0.3$, so we extract six quantities (DegDistP1..DegDistP6 percentiles) from any degree distributions. If we increase the value of $K$, we should normally decrease the value of $\sigma$ so that most of the interval points stay in the range of existing node degrees. Smaller values for $\sigma$ also necessitate larger values for $K$. Large values (e.g., $K=100$) and small values (e.g., $K=1$) for $K$ will also decrease the distinction power of the extracted features vector. The specified values for $\sigma$ and $K$ are found through trial and error. Equation \ref{eq:IntervalPoint} shows the interval points of the degree distribution and Equation \ref{eq:DegDistP} specifies the probability for a node degree to sit in the $ i $th interval. The set of six percentiles (DegDistP1..DegDistP6) are used as the network features representing the degree distribution.\\
Let $IP_{i}$ be the $i$th interval point and $D$ be the degree random variable.
\end{itemize}

\begin{eqnarray}
	\label{eq:IntervalPoint}
	IP_{i} = \left\{
		\begin{matrix}
			min(D) &&i=1\\ 
			\mu-(\frac{k}{2}-i+1)p\sigma &&i=2..K\\
			max(D) &&i=K+1
		\end{matrix}\right.
\end{eqnarray}

\begin{eqnarray}
	\label{eq:DegDistP}
	DegDistP_{i} = P(IP_{i} \: < D \: < IP_{i+1}), \: i=1..K
\end{eqnarray}

\subsection{Learning the Classifier}
The third step of the proposed methodology is the utilization of a supervised machine learning algorithm. The learning algorithm constructs the network classifier based on the features of generated network instances as the training data. Each record of the training data consists of the structural features -as described in the previous subsection- of a generated network along with the label of its generative model. By the means of supervised algorithms, we can learn from this training data a classifier which predicts the best generative model for a given network with the specified structural features.\\

We examined several supervised learning algorithms such as decision tree learning \cite{QC4.5, MulticlassADT}, Bayesian networks \cite{BayesianRef}, support vector machines \cite{SMORef} (SVM) and neural networks \cite{NeuralNetRef} among which the LADTree method showed better results. A short description of examined learning algorithms is presented in Appendix \ref{App:LearningAlg}. In our experiments, although some methods (such as Bayesian networks) resulted in a small improvement in the accuracy of the learned classifier, but the decision tree learned by LADTree algorithm was obviously more robust and less sensitive to noises than other learning methods. The robustness to noise analysis is described in the evaluation section. To avoid over-fitting, we always used stratified 10-fold cross-validation.\\

\subsection{Network Models}
Among several existing network generative models, we have selected seven important models: Kronecker Graphs \cite{Kronecker} Model, Forest Fire \cite{GraphsOverTime} Model , Random Typing Generator \cite{RTG} Model, Preferential Attachment \cite{EmergenceOfScaling} Model, Small World \cite{SmallWorld} Model, Erd{\"o}s-R\'enyi \cite{ER} Model and Random Power Law \cite{RP} Model. The selected models are the state of the art methods of network generation. The existing model selection methods such as Ref. \cite{ModelSelection} and Ref. \cite{Drosophila} have ignored some new and important generative models such as Kronecker Graphs \cite{Kronecker}, Forest Fire \cite{GraphsOverTime} and RTG \cite{RTG} models.

\section{\label{Evaluation}Evaluation}
In this section, we evaluate our proposed method of model selection (GMSCN). We also compare GMSCN with the baseline method \cite{ModelSelection} and show that it outperforms state of the art methods with respect to different criteria.\\

Despite most of the existing methods, GMSCN has no dependency on the size of the given network. In other words, we ignore the number of nodes of the target network and we only consider its density in generating the training data. Because the baseline method is dependent on the size of the target network, we evaluate the methods in two stages. In the first stage, we fix the size of the generated networks to prepare a fair condition for comparing GMSCN with the baseline method. Although size-dependence is a drawback for the baseline method, the evaluation shows that GMSCN outperforms the baseline method even in fixed network size condition. In the second stage, we allow the generation models to synthesize networks of different sizes. In this stage, we show that the size diversity of generated networks does not affect the accuracy of the learned decision tree.
As described in Section \ref{ProposedMethod}, GMSCN is based on learning a decision tree from a training set of generated networks. In each evaluation stage, we generated 100 networks from each network generative model and with seven candidate models, we gathered 700 generated networks. We used these network instances as the training and test data for learning the decision tree. 

\subsection{The Baseline method}
\begin{figure}
\includegraphics{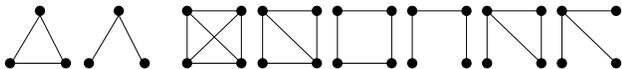}
\caption{\label{fig:Graphlet}The graphlets with three and four nodes}
\end{figure}

We have selected the graphlet-based method proposed by Janssen et al. \cite{ModelSelection} as the baseline method. The baseline method has some similarities to GMSCN: it is based on considering some network generative models and then learning a decision tree for network classification with the aid of a set of generated networks. In the baseline method, eight graphlet counts are considered as the network features. All subgraphs with three nodes (two graphlets) and four nodes (six graphlets) are considered in the baseline method (Fig. \ref{fig:Graphlet}). A similar approach is also proposed by Middendorf et al. \cite{Drosophila}, with distinctions on the learning algorithm and the set of candidate generative models.
The graphlet-based method is selected as the baseline because it is a new method and its evaluations show a high accuracy, and it is proposed similarly in different research domains such as social networks \cite{ModelSelection} and protein networks \cite{Drosophila}.\\

Despite the similarities, there exist some important differences between GMSCN and the baseline method. First, the baseline method is based on counting graphlets in networks while GMSCN proposes a wider set of local and global features. Janssen et al. \cite{ModelSelection} conclude that considering structural features does not improve the accuracy of the graphlet-based classifier, but we will show that choosing a better set of local and global network features and with the aid of our proposed degree distribution quantification method, structural features will play an undeniable role in model selection. Second, the baseline method is size-dependent, i.e., it considers both the size and the density of the target network, and it generates network instances according to these two properties. On the other hand, GMSCN is size-independent and we only consider the density of the target network in the network generation phase. Third, GMSCN employs newer and more important generative models such as the Kronecker Graphs \cite{Kronecker} model, the Forest Fire \cite{GraphsOverTime} model and the RTG \cite{RTG} model. Fourth, we examined different learning algorithms and then selected LADTree as the best learning algorithm for this application. Our evaluation of GMSCN is more thorough, considering different evaluation criteria. We have also presented a new algorithm for quantifying the network degree distribution.
\\
Graphlet counting is a very time consuming task and there is no efficient algorithm for computing the full counts of graphlets for large networks. To handle the algorithmic complexity, most of the graphlet-counting methods (e.g., Refs. \onlinecite{ModelSelection, NetMotifDiscovery, ModelsInBiology, CellBiology}) propose a sampling phase before counting the graphlets. But the sampling algorithm may affect the graphlet counts and the resulting counts may be biased towards the features of the sampling algorithm. It is also possible to estimate the graphlet counts with approximate algorithms \cite{SubgraphCounting, Graft}, but this approach may also bring remarkable errors in graphlet counts. To prepare a fair comparison situation, we have counted the exact number of graphlets in original networks and have not employed any sampling or approximation algorithms.
It is worth noting that reported accuracy of the baseline method in this paper is different from the report of the original paper \cite{ModelSelection}, mainly because the set of generative models are not the same in the two papers.

\subsection{\label{subsec:AccuracyOfTheModelClassifier}Accuracy of the Model Classifier}
We first set a fixed size for generated networks of the dataset and generate networks with about 4096 nodes. Almost all the generated networks in our dataset contain 4096 nodes, but the networks generated by RTG \cite{RTG} model have small variations in their size. Number of nodes in these networks is in the range of 4000 to 4200 and this is because the exact number of nodes is not configurable in the RTG model. Since the Kronecker Graphs model generates networks with $2^{x}$ nodes in its original form, we chosen 4096 $(2^{12})$ as the size of the networks. The average density of networks in this dataset is equal to 0.0024.\\

\begin{figure}
\includegraphics{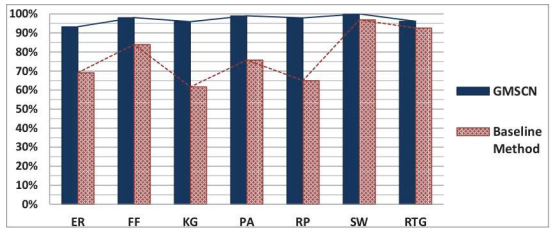}
\caption{\label{fig:TwoMethodsPrecisions}Precision of GMSCN compared to baseline method for different generative models}
\end{figure}

\begin{figure}
\includegraphics{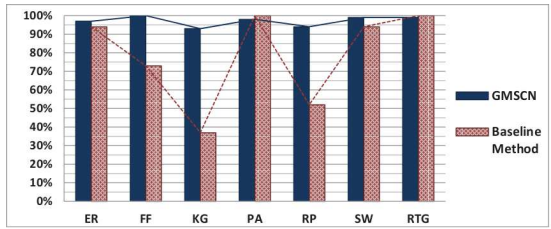}
\caption{\label{fig:TwoMethodsRecalls}Recall of GMSCN compared to baseline method for different generative models}
\end{figure}

In addition to overall accuracy, we evaluate the precision and recall of the learned decision tree for different network models. ``Precision" shows the percentage of correctly classified instances calculated for each category (e.g., $ Precision_{FF} = \frac{number \: of \: correctly \: predicted \: FF \: intances}{number \: of \: FF \: predicted \: instances}$), ``Recall" illustrates the ability of the method in finding the instances of a category (e.g., $ Recall_{FF} = \frac{number \: of \: correctly \: predicted \: FF \: intances}{number \: of \: FF \: instances}$), and ``Accuracy" is an indicator of overall effectiveness of the classifier across the entire dataset (i.e., $ Accuracy = \frac{number \: of \: correctly \: predicted \: intances}{total \: number \: of \: instances}$). The overall accuracy of GMSCN is 97.14\% while the accuracy of the baseline method is 78.57\% which indicates 18.57\% improvements. Fig. \ref{fig:TwoMethodsPrecisions} and Fig. \ref{fig:TwoMethodsRecalls} show the precision and recall of GMSCN and the baseline method respectively for different network models. In addition to an apparent improvement in the precision and recall for most of the generative models, the figures show the stability (less undesired deviation) of GMSCN over the baseline method. The accuracy and precision of GMSCN show small deviation for different generative models, while these measures for baseline method vary in a wide range. Table \ref{tab:GMSCNResults} shows the details of GMSCN results for different network models. For example, the first row of this table indicates that among 700 network instances, 104 networks are predicted to be generated by the ER model but in fact 97 (out of 104) instances are ER, six instances are the KG model and one is generated by the SW model. Because we have utilized cross-validation, all of the 700 network instances are included in the evaluation. Table \ref{tab:BaselineResults} shows corresponding results for the baseline method.\\

It is worth noting that considering both the graphlet counts and the structural features does not improve the accuracy of the classifier considerably. Since we want to prepare a size-independent and efficient method, we do not consider the graphlet counts in feature vectors. 

\begin{table*}[htbp]
  \centering
  \caption{\label{tab:GMSCNResults}Precision, Recall and Accuracy of GMSCN for different generative models}
 	\begin{ruledtabular}
    \begin{tabular}{lrrrrrrrr}
    \toprule
          & \multicolumn{1}{l}{True ER} & \multicolumn{1}{l}{True FF} & \multicolumn{1}{l}{True KG} & \multicolumn{1}{l}{True PA} & \multicolumn{1}{l}{True RP} & \multicolumn{1}{l}{True SW} & \multicolumn{1}{l}{True RTG} & \multicolumn{1}{l}{\textbf{Class Precision}} \\ \hline
    \midrule
    Pred. ER & \multicolumn{1}{c}{97} & \multicolumn{1}{c}{0} & \multicolumn{1}{c}{6} & \multicolumn{1}{c}{0} & \multicolumn{1}{c}{0} & \multicolumn{1}{c}{1} & \multicolumn{1}{c}{0} & \multicolumn{1}{c}{93.27\%} \\
    Pred. FF & \multicolumn{1}{c}{0} & \multicolumn{1}{c}{100} & \multicolumn{1}{c}{0} & \multicolumn{1}{c}{0} & \multicolumn{1}{c}{2} & \multicolumn{1}{c}{0} & \multicolumn{1}{c}{0} & \multicolumn{1}{c}{98.04\%} \\
    Pred. KG & \multicolumn{1}{c}{2} & \multicolumn{1}{c}{0} & \multicolumn{1}{c}{93} & \multicolumn{1}{c}{2} & \multicolumn{1}{c}{0} & \multicolumn{1}{c}{0} & \multicolumn{1}{c}{0} & \multicolumn{1}{c}{95.88\%} \\
    Pred. PA & \multicolumn{1}{c}{1} & \multicolumn{1}{c}{0} & \multicolumn{1}{c}{0} & \multicolumn{1}{c}{98} & \multicolumn{1}{c}{0} & \multicolumn{1}{c}{0} & \multicolumn{1}{c}{0} & \multicolumn{1}{c}{98.99\%} \\
    Pred. RP & \multicolumn{1}{c}{0} & \multicolumn{1}{c}{0} & \multicolumn{1}{c}{1} & \multicolumn{1}{c}{0} & \multicolumn{1}{c}{94} & \multicolumn{1}{c}{0} & \multicolumn{1}{c}{1} & \multicolumn{1}{c}{97.92\%} \\
    Pred. SW & \multicolumn{1}{c}{0} & \multicolumn{1}{c}{0} & \multicolumn{1}{c}{0} & \multicolumn{1}{c}{0} & \multicolumn{1}{c}{0} & \multicolumn{1}{c}{99} & \multicolumn{1}{c}{0} & \multicolumn{1}{c}{100.00\%} \\
    Pred. RTG & \multicolumn{1}{c}{0} & \multicolumn{1}{c}{0} & \multicolumn{1}{c}{0} & \multicolumn{1}{c}{0} & \multicolumn{1}{c}{4} & \multicolumn{1}{c}{0} & \multicolumn{1}{c}{99} & \multicolumn{1}{c}{96.12\%} \\
    \textbf{Class Recall} & \multicolumn{1}{c}{97\%} & \multicolumn{1}{c}{100\%} & \multicolumn{1}{c}{93\%} & \multicolumn{1}{c}{98\%} & \multicolumn{1}{c}{94\%} & \multicolumn{1}{c}{99\%} & \multicolumn{1}{c}{99\%} & \multicolumn{1}{l}{\textbf{Accuracy: 97.14\% }} \\
    \bottomrule
    \end{tabular}
    \end{ruledtabular}
\end{table*}

\begin{table*}[htbp]
  \centering
  \caption{\label{tab:BaselineResults}Precision, Recall and Accuracy of the baseline method for different generative models}
	\begin{ruledtabular}
    \begin{tabular}{lrrrrrrrr}
    \toprule
          & \multicolumn{1}{l}{True ER} & \multicolumn{1}{l}{True FF} & \multicolumn{1}{l}{True KG} & \multicolumn{1}{l}{True PA} & \multicolumn{1}{l}{True RP} & \multicolumn{1}{l}{True SW} & \multicolumn{1}{l}{True RTG} & \multicolumn{1}{l}{\textbf{Class Precision}} \\ \hline
    \midrule
    Pred. ER & \multicolumn{1}{c}{94} & \multicolumn{1}{c}{1} & \multicolumn{1}{c}{30} & \multicolumn{1}{c}{0} & \multicolumn{1}{c}{11} & \multicolumn{1}{c}{0} & \multicolumn{1}{c}{0} & \multicolumn{1}{c}{69.12\%} \\
    Pred. FF & \multicolumn{1}{c}{0} & \multicolumn{1}{c}{73} & \multicolumn{1}{c}{2} & \multicolumn{1}{c}{0} & \multicolumn{1}{c}{6} & \multicolumn{1}{c}{6} & \multicolumn{1}{c}{0} & \multicolumn{1}{c}{83.91\%} \\
    Pred. KG & \multicolumn{1}{c}{6} & \multicolumn{1}{c}{0} & \multicolumn{1}{c}{37} & \multicolumn{1}{c}{0} & \multicolumn{1}{c}{17} & \multicolumn{1}{c}{0} & \multicolumn{1}{c}{0} & \multicolumn{1}{c}{61.67\%} \\
    Pred. PA & \multicolumn{1}{c}{0} & \multicolumn{1}{c}{0} & \multicolumn{1}{c}{26} & \multicolumn{1}{c}{100} & \multicolumn{1}{c}{6} & \multicolumn{1}{c}{0} & \multicolumn{1}{c}{0} & \multicolumn{1}{c}{75.76\%} \\
    Pred. RP & \multicolumn{1}{c}{0} & \multicolumn{1}{c}{23} & \multicolumn{1}{c}{5} & \multicolumn{1}{c}{0} & \multicolumn{1}{c}{52} & \multicolumn{1}{c}{0} & \multicolumn{1}{c}{0} & \multicolumn{1}{c}{65.00\%} \\
    Pred. SW & \multicolumn{1}{c}{0} & \multicolumn{1}{c}{3} & \multicolumn{1}{c}{0} & \multicolumn{1}{c}{0} & \multicolumn{1}{c}{0} & \multicolumn{1}{c}{94} & \multicolumn{1}{c}{0} & \multicolumn{1}{c}{96.91\%} \\
    Pred. RTG & \multicolumn{1}{c}{0} & \multicolumn{1}{c}{0} & \multicolumn{1}{c}{0} & \multicolumn{1}{c}{0} & \multicolumn{1}{c}{8} & \multicolumn{1}{c}{0} & \multicolumn{1}{c}{100} & \multicolumn{1}{c}{92.59\%} \\
    \textbf{Class Recall} & \multicolumn{1}{c}{94\%} & \multicolumn{1}{c}{73\%} & \multicolumn{1}{c}{37\%} & \multicolumn{1}{c}{100\%} & \multicolumn{1}{c}{52\%} & \multicolumn{1}{c}{94\%} & \multicolumn{1}{c}{100\%} & \multicolumn{1}{l}{\textbf{Accuracy: 78.57\%}} \\
    \bottomrule
    \end{tabular}
    \end{ruledtabular}
\end{table*}

\subsection{\label{subsec:SizeIndependence}Size Independence}
GMSCN for model selection is independent from the size of the target network. When we want to find the best model fitting a real network, we can discard the number of nodes in the network and generate the training data only according to its density. The size-independence is an important feature of GMSCN which is missing in the baseline method. This feature is especially important when we want to find the generative model for a relatively large network. In this condition, we can generate the training network instances with smaller sizes than the target network. This feature also increases the applicability, scalability and performance of GMSCN.\\

For evaluating the dependency of GMSCN to the size of the network, we generate a new dataset with networks of different sizes. Instead of fixing the number of nodes in each network instance (such as about 4096 nodes in the previous evaluation) we allow networks with different number of nodes in the dataset. In this test, with each of the generative models, we generated 100 networks of different sizes: 24 networks with 4,096 nodes, 24 networks with 32,768 nodes, 24 networks with 131,072 nodes, 24 networks with 524,288 nodes and four networks with 1,048,576 nodes. Again, the only exception is the RTG model which generates networks with small variations from the specified sizes. The node counts are powers of two because the original version of Kronecker graph model is able to generate networks with $2^{n}$ nodes. The average density of networks in this dataset is equal to 0.000885.\\

\begin{figure}
\includegraphics{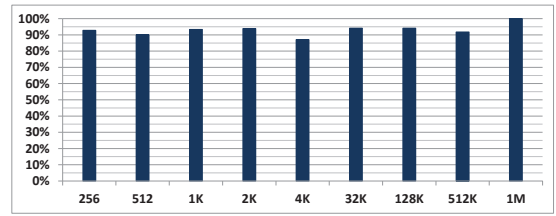}
\caption{\label{fig:SizeIndependence}Accuracy of GMSCN for different network sizes.}
\end{figure}

Table \ref{tab:SizeIndependent} shows the precision and recall of GMSCN for this dataset. In this evaluation, the overall accuracy of the classifier is 97.29\% which is very close to the accuracy of the system in the evaluation with fixed network sizes. This fact shows that GMSCN is not dependent on the size of the target network. The average density of networks in this dataset (0.000885) is different from the average density of networks in the fixed-size dataset (0.0024). So, the model selection is also performing well for different densities of the given network. We also extended this experiment to ensure that there is no meaningful lower bound for GMSCN in terms of network size. The new experiment is configured similar to the previous trial, but it examines a wider range of network sizes. Fig. \ref{fig:SizeIndependence} plots the result of this experiment at each number of nodes. It indicates that GMSCN shows good performance for the varying network sizes. Obviously, the baseline method is size-dependent \cite{ModelSelection, Drosophila} because the graphlet counts completely depend on the  size of the network. So, it is not necessary to show the precision and recall of the baseline method for dataset of networks of different sizes. We ignored such a useless evaluation because the calculation of graphlet counts for large networks is very time consuming.

\begin{table}[htbp]
  \centering
  \caption{\label{tab:SizeIndependent}Precision and Recall of GMSCN for networks of different sizes}
	\begin{ruledtabular}
    \begin{tabular}{lrrrrrrr}
    \toprule
     & \multicolumn{1}{l}{ER} & \multicolumn{1}{l}{FF} & \multicolumn{1}{l}{KG} & \multicolumn{1}{l}{PA} & \multicolumn{1}{l}{RP} & \multicolumn{1}{l}{SW} & \multicolumn{1}{l}{RTG} \\ \hline
    \midrule
    \textbf{Precision} & \multicolumn{1}{c}{96.0\%} & \multicolumn{1}{c}{98.0\%} & \multicolumn{1}{c}{95.0\%} & \multicolumn{1}{c}{98.0\%} & \multicolumn{1}{c}{97.9\%} & \multicolumn{1}{c}{100\%} & \multicolumn{1}{c}{96.1\%} \\
    \textbf{Recall} & \multicolumn{1}{c}{96.0\%} & \multicolumn{1}{c}{100\%} & \multicolumn{1}{c}{95.0\%} & \multicolumn{1}{c}{99.0\%} & \multicolumn{1}{c}{94.0\%} & \multicolumn{1}{c}{99.0\%} & \multicolumn{1}{c}{98.0\%} \\
    \bottomrule
    \end{tabular}
    \end{ruledtabular}
\end{table}

\subsection{Robustness to Noise}
We also evaluate the robustness of GMSCN with respect to random changes in networks. For each test-case network, we randomly select a fraction of edges, rewire them to random nodes, and test the accuracy of the classifier for the resulting network. We start from the pure network samples and in each step, we change five percent of the edges until all the edges (100 percent change) are randomly rewired. In other words, in addition to pure networks, we generated 20 test-sets with from zero to 100 percent edge changes, each of which containing 700 network samples from seven generative models.\\

\begin{figure}
\includegraphics{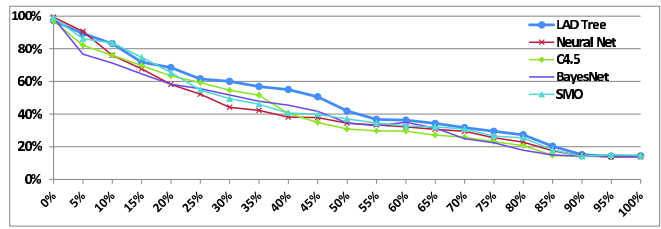}
\caption{\label{fig:Robustness}Robustness of the different classification methods with respect to random edge rewiring.}
\end{figure}

As discussed before, we have chosen LADTree as the supervised learning algorithm in GMSCN. Fig. \ref{fig:Robustness} shows the average accuracy of GMSCN for different random change fractions. This figure shows the effect of choosing different learning algorithms for GMSCN. As the figure shows, LADTree results in a more robust classifier for this application, since it is less sensitive to noise. The accuracy of GMSCN is smoothly decreasing nearly linear with random changes. There is no sudden drop in the chart of the GMSCN (based on LADTree). With 100 percent random changes (the right end of the diagram), the accuracy of the classifier reaches the value of 14.43 percent, which is near to 1/7 (i.e., $ \frac{1}{number \: of \: candidate \: models} $). This is due to existence of seven network models and indicates that almost all the characteristics of the generative model is eliminated from a generated network with 100 percent edge rewiring. \\

\subsection{Scalability and Performance}
The aim of GMSCN is finding a generative model best fitting a given real network. We define the scalability of such a method as its ability to handle networks of large sizes as the input. Noting to the methodology of the proposed method (Fig. \ref{fig:Methodology}), the most time-consuming part of the model classification is the feature extraction task. For the feature extraction task, GMSCN is obviously more scalable than the baseline method. There is no efficient algorithm for counting the graphlets in large networks. The selected network features in GMSCN (effective diameter, clustering coefficient, transitivity, assorativity and degree distribution percentiles) are efficiently computable by existing algorithms. We have also discarded ``timely to extract" features such as ``average path length" because their extraction has more computationally complex algorithms.
\\
Most of the graphlet-based methods such as Ref. \cite{ModelSelection} and Ref. \cite{Drosophila} try to increase their scalability by incorporating a pre-stage of network sampling with very small rates such as 0.01\% (one out of 10,000) in Ref. \cite{ModelSelection}. But such sampling rates decreases the accuracy of graph counts and the chosen sampling algorithm will also bias the graph counts. On the other hand, if sampling or approximation algorithms are accepted for baseline method, these techniques will improve the performance of GMSCN too. In other words, utilization of sampling and approximation algorithms increases the scalability of both of the baseline method and GMSCN similarly. Some notes about the implementation and evaluation of GMSCN are presented in the Appendix \ref{App:ImplementationNotes}.

\subsection{Effectiveness of the Degree Distribution Quantification Method}
As described in Section \ref{ProposedMethod}, we have proposed a new method for the quantification of the degree distribution based on its mean and standard deviation. In this subsection, we test the effectiveness of this quantification method. We show that without the proposed features of degree distribution, the accuracy of the network classifier will diminish. Table \ref{tab:WithoutDegDist} shows the results of GMSCN by eliminating six features related to the degree distribution (DegDistP1..DegDistP6 percentiles). By this change, the overall accuracy of the method decreases about eight percent (from 97.14\% to 89.29\%). This can be seen by comparing the values in Table \ref{tab:WithoutDegDist} with those of Table \ref{tab:GMSCNResults} which reflects the results of GMSCN when employing all the features. Precision and recall are improved for almost all the models with incorporating features related to the degree distribution. This fact shows the effectiveness of our proposed quantification method for degree distribution.

\begin{table}[htbp]
  \centering
  \caption{\label{tab:WithoutDegDist}The results of GMSCN after excluding the features of degree distribution}
	\begin{ruledtabular}
    \begin{tabular}{lrrrrrrr}
    \toprule
     & \multicolumn{1}{l}{ER} & \multicolumn{1}{l}{FF} & \multicolumn{1}{l}{KG} & \multicolumn{1}{l}{PA} & \multicolumn{1}{l}{RP} & \multicolumn{1}{l}{SW} & \multicolumn{1}{l}{RTG} \\ \hline
    \midrule
    \textbf{Precision} & 89.1\% & 100\% & 79.41\% & 94.17\% & 76.47\% & 96\% & 90.7\% \\
    \textbf{Recall} & 90\% & 95\% & 81\% & 97\% & 78\% & 96\% & 88\% \\
    \bottomrule
    \end{tabular}
    \end{ruledtabular}  
\end{table}

\section{\label{CaseStudy}Case study}
We applied GMSCN for some real networks. The real network instances and the result of applying GMSCN on these networks are illustrated here:
\begin{enumerate}
\item ``dblp\_cite"  \footnote{http://dblp.uni-trier.de/xml} (with 475,886 nodes and 2,284,694 edges) is a network which is extracted from the DBLP service. This network shows the citation network among scientific papers. GMSCN proposes Forest Fire as the best fitting generative model for this network. Leskovec et al. \cite{GraphsOverTime} also propose Forest Fire model for two similar graphs of arXiv and patent citation networks.

\item	``dblp\_collab" \footnote{http://dblp.uni-trier.de/xml} (with 975,044 nodes and 3,489,572 edges) is a co-authorship network of papers indexed in the DBLP service. A node in this network represents an author and an edge indicates at least one collaboration in writing papers between the two authors. GMSCN suggests Forest Fire for this network instance too. 

\item	``p2p-Gnutella08" \footnote{http://snap.stanford.edu} (with 6,301 nodes and 20,777 edges) is a relatively small P2P network with about 6000 nodes. The best fitting model suggested by GMSCN for this network instance is Kronecker Graphs. 

\item	Slashdot, as a technology-related news website, presented the Slashdot Zoo which allowed users to tag each other as friends. ``Slashdot0902" \footnote{http://snap.stanford.edu} (with 82,168 nodes and 543,381 edges) is a network of friendship links between the users of Slashdot, obtained in February 2009. The output of GMSCN for this social network is the Random Power Law model. 

\item	In the ``web-Google" \footnote{http://snap.stanford.edu} (with 875,713 nodes and 4,322,051 edges) network, the nodes represent web pages and directed edges represent hyperlinks among them. We ignored the direction of the links and considered the network as a simple undirected graph. The random Power Law model is also proposed for this network by GMSCN. 

\item	``Email-EuAll" \footnote{http://konect.uni-koblenz.de} (with 265,214 nodes and 365,025 edges) is a communication network of email contacts which is predicted to follow the RTG model. 

\item	Finally, for the small network of ``Email-URV" \footnote{http://deim.urv.cat/\textasciitilde aarenas} (with 1,133 nodes and 5,451
 edges), which is another communication network of emails, GMSCN suggests the Small World model.
\\
\end{enumerate}
As explained above, various real networks, which are selected from a wide range of sizes, densities, and domains, are categorized in different network models by the GMSCN classifier. This fact indicates that no generative model is dominated in GMSCN for real networks and it suggests different models for different network structures. The case study also verifies that no generative model is sufficient for synthesizing networks similar to real networks and we should find the best model fitting to the target network in each application. As a result, it is worth noting that the task of generative model selection is an important stage before generating network instances.

\section{\label{Discussion}Discussion}
We evaluated GMSCN from different perspectives. GMSCN proposes a size-independent methodology for building the network classifier based on a wide range of local and global network features as the inputs of a decision tree. It shows a high accuracy in predicting the generative model for a given network. It is tolerant and insensitive to small network changes. In addition to size-independence, GMSCN outperforms the baseline method –that only considers local features of graphlet counts with respect to accuracy and efficiency. A new structural feature is also proposed in GMSCN which quantifies the network degree distribution.
\\
One may argue that the size of the training set (700 network instances) is relatively small for a machine learning task. But we have actually utilized many more network instances in the process of evaluating GMSCN. Our dataset for evaluating GMSCN includes 15,400 different network instances: 700 instances in the fixed-size evaluation, 700 instances in the size-independence test and 14,000 (20$\times$700) instances in the robustness test. The dataset size seems to be sufficient for evaluating the learned classifier because the network instances are generated with different parameters (e.g., different sizes) and the results for various evaluation steps are stable.
\\
It can also be argued that the definition of ``Accuracy" in the evaluations is not fair. When we compute the accuracy of the classifier, given that a network is generated precisely according to one of seven models, the classifier attempts to determine the generative model. One may argue that real networks are unlikely to be determined by one of these models, so accuracy in predicting the origin of artificially generated networks does not necessarily imply accuracy for real networks. But we have shown that GMSCN is able to classify synthesized network instances even with random noises (in subsection \ref{subsec:SizeIndependence}). In other words, networks that are not completely compatible to one of the generative models are also well categorized with GMSCN. We should note that no accepted benchmark exists for suggesting the best generative model for real networks. So, the computation of the actual accuracy of a model selection algorithm for real networks is fundamentally impossible.
\\
Considering the existing model selection methods, we summarize the main distinctions and contributions of GMSCN here. First, we have proposed new structural network features based on the quantification of degree distribution. We have shown the effectiveness of these features in improving the accuracy of the model selection method. Second, we proposed a set of local and global network features for the problem of model selection. The baseline method suggests a set of graphlet counts that are limited local features and the evaluations show that such features are not sufficient for this application. It is not possible to capture important characteristics of real networks such as heavy-tailed degree distribution, small path lengths, and degree correlation (assortativity) only by counting graphlets, while such characteristics are among the main distinctions of generative models. For example, the Small World model generates networks with high clustering and small path lengths and artificial networks generated by most of the models demonstrate heavy-tailed degree distributions. Third, GMSCN is a size-independent method and the learned classifier is applicable for networks of different sizes. This is an important feature especially in the case of suggesting a generative model for a large network. In this case, we can generate the training set of artificial networks with a relatively smaller number of nodes. Fourth, although our proposed methodology is not dependent on the generative models, we have chosen seven important and outstanding network generative models as the candidate models of the classifier. Important models such as Kronecker graphs, Forest Fire and RTG are not considered in similar existing methods. Fifth, we have investigated different learning algorithms and reached LADTree as the most robust learning algorithm for this application. Sixth, we have presented a diverse set of evaluations for GMSCN with different criteria such as precision, recall, accuracy, robustness to noise, size-independence, scalability and effectiveness of the features.

\section{\label{Conclusion}Conclusion}
In this paper, we proposed a new method (GMSCN) for network model selection. This method, which is based on learning a decision tree, finds the best model for generating complex networks similar to a specified network instance. The structural features of the given network instance are utilized as the input of the decision tree and the result is the best fitting model. GMSCN outperforms the existing methods with respect to different criteria. The accuracy of GMSCN shows a considerable improvement over the baseline method \cite{ModelSelection}. In addition, the set of supported generative models in GMSCN contains wider, newer and more important generative models such as Kronecker graphs, Forest Fire and RTG. Despite most of the existing methods, GMSCN is independent from the size of the input network. GMSCN is a robust model and insensitive to small network changes and noises. It is also a scalable method and its performance is obviously better than the baseline method. GMSCN also includes a new and effective algorithm for the quantification of network degree distribution. We have examined different learning algorithms and as a result, decision tree learning by LADTree method was the most accurate and robust model. We showed that the local structural features, such as graphlet counts, are insufficient for inferring the network mechanisms and it is a must to consider a wider range of local and global structural features to be able to predict the network growth mechanisms.\\

In future, we will investigate the effect of network structural features and growth mechanisms on dynamics and behavior of the network when it is faced with different processes. For example, we will evaluate the similarity of the information diffusion process in a network and its counterparts synthesized by the selected network generation model.

\begin{acknowledgments}
We wish to thank Masoud Asadpour, Mehdi Jalili and Abbas Heydarnoori for their great comments.
\end{acknowledgments}

\appendix

\section{\label{App:LearningAlg}Brief Introduction to Classification Methods}
Machine Learning is a subfield of Artificial Intelligence in which the main goal is to learn knowledge through experience. Classification is a learning task of inferring a classification function from labeled training data. Here, we explain some classifiers that are used in this paper.
\\
\textit{\textbf{Support Vector Machines (SVM)}}\cite{SMORef}. SVM performs a classification by mapping the inputs into a high-dimensional feature space and constructing hyperplanes to categorize the data instances. The best hyperplanes are those that cause the largest margin among the classes. The parameters of such a maximum-margin hyperplane are derived by solving an optimization problem. Sequential Minimal Optimization (SMO) \cite{SMORef} is a common method for solving the optimization problem.
\\
\textit{\textbf{Bayesian Networks Learning}} \cite{BayesianRef}. A Bayesian network model is a probabilistic graphical model that represents a set of random variables and their conditional dependencies by a directed acyclic graph. The nodes in this graph represent the random variables and an edge shows a conditional dependency between two variables. Bayesian network learning aims to create a network that “best describes” the probability distribution over the training data. To find the best network among the set of possible Bayesian networks, the heuristic search techniques has been frequently used in the literature.
\\
\textit{\textbf{Artificial Neural Networks}}\cite{NeuralNetRef}. ANN is inspired by human brain neural network. An ANN consists of neuron units, arranged in layers and connected with weighted links, which convert an input vector into some outputs.  Usually, the networks are defined to be feed-forward, with no feedback to the previous layer.  In the training phase, the weights of the links are tuned to adapt an ANN to the training data. Back-propagation algorithm is a common method for the training phase.
\\
\textit{\textbf{C4.5 Decision Tree Learning}}\cite{QC4.5}. A decision tree is a tree structure of decision rules which can be used as a classification function (leaf nodes show the returned classes). C4.5 constructs a decision tree based on a labeled training data. C4.5 uses “information entropy” to evaluate the goodness of branches in the tree.
\\
\textit{\textbf{LADTree}}\cite{MulticlassADT}.This classifier generates a multi-class alternating decision tree and it uses the “boosting” strategy. Boosting is a well-established classification technique that combines some weak classifiers to form a single powerful classifier. A prediction node in a LADTree includes a score for each of candidate classes. LADTree calculates confidences for different classes according to their visited score in prediction nodes, and it returns the best class according to the confidences.

\section{\label{App:ImplementationNotes}Implementation Notes}
To implement Kronecker Graphs, Forest Fire model, Preferential Attachment, Small World, and Random Power Law models, we utilized the SNAP  library (\url{http://snap.stanford.edu/snap/}). The implementation of RTG model is available in a MATLAB library (\url{http://www.cs.cmu.edu/~lakoglu/}). We also developed our own implementation of the ER model. The features are extracted by the aid of different network analysis tools. The igraph package (\url{http://igraph.sourceforge.net/}) of the R project  helped us calculate the assortativity and transitivity measures. We used the SNAP library for measuring the effective diameter, average clustering coefficient, density and also the graphlet counts. Since we proposed a new method for quantifying network degree distribution, we have implemented this method ourselves. We utilized RapidMiner  as an open source tool for machine learning. The implementation of LADTree and Bayesian network learning and SVM are actually part of the Weka  tool which is embedded in RapidMiner. The amount of computation needed for this research, especially counting the exact number of graphlets, was enormous. We utilized three virtual machines on a super-computer for this enormous computation task, each of which simulated a computer with 16 processing cores of 2.8 GHz and 24 GB of memory. Most of the computation time was spent for counting the graphlets of the generated network instances.

\bibliography{shortbib}

\begin{thebibliography}{59}%
\makeatletter
\providecommand \@ifxundefined [1]{%
 \@ifx{#1\undefined}
}%
\providecommand \@ifnum [1]{%
 \ifnum #1\expandafter \@firstoftwo
 \else \expandafter \@secondoftwo
 \fi
}%
\providecommand \@ifx [1]{%
 \ifx #1\expandafter \@firstoftwo
 \else \expandafter \@secondoftwo
 \fi
}%
\providecommand \natexlab [1]{#1}%
\providecommand \enquote  [1]{``#1''}%
\providecommand \bibnamefont  [1]{#1}%
\providecommand \bibfnamefont [1]{#1}%
\providecommand \citenamefont [1]{#1}%
\providecommand \href@noop [0]{\@secondoftwo}%
\providecommand \href [0]{\begingroup \@sanitize@url \@href}%
\providecommand \@href[1]{\@@startlink{#1}\@@href}%
\providecommand \@@href[1]{\endgroup#1\@@endlink}%
\providecommand \@sanitize@url [0]{\catcode `\\12\catcode `\$12\catcode
  `\&12\catcode `\#12\catcode `\^12\catcode `\_12\catcode `\%12\relax}%
\providecommand \@@startlink[1]{}%
\providecommand \@@endlink[0]{}%
\providecommand \url  [0]{\begingroup\@sanitize@url \@url }%
\providecommand \@url [1]{\endgroup\@href {#1}{\urlprefix }}%
\providecommand \urlprefix  [0]{URL }%
\providecommand \Eprint [0]{\href }%
\providecommand \doibase [0]{http://dx.doi.org/}%
\providecommand \selectlanguage [0]{\@gobble}%
\providecommand \bibinfo  [0]{\@secondoftwo}%
\providecommand \bibfield  [0]{\@secondoftwo}%
\providecommand \translation [1]{[#1]}%
\providecommand \BibitemOpen [0]{}%
\providecommand \bibitemStop [0]{}%
\providecommand \bibitemNoStop [0]{.\EOS\space}%
\providecommand \EOS [0]{\spacefactor3000\relax}%
\providecommand \BibitemShut  [1]{\csname bibitem#1\endcsname}%
\let\auto@bib@innerbib\@empty
\bibitem [{\citenamefont {Newman}(2003)}]{Newman}%
  \BibitemOpen
  \bibfield  {author} {\bibinfo {author} {\bibfnamefont {M.~E.}\ \bibnamefont
  {Newman}},\ }\bibfield  {title} {\enquote {\bibinfo {title} {The structure
  and function of complex networks},}\ }\href@noop {} {\bibfield  {journal}
  {\bibinfo  {journal} {SIAM review}\ }\textbf {\bibinfo {volume} {45}},\
  \bibinfo {pages} {167--256} (\bibinfo {year} {2003})}\BibitemShut {NoStop}%
\bibitem [{\citenamefont {Albert}\ and\ \citenamefont
  {Barab{\'a}si}(2002)}]{StatisticalMechanics}%
  \BibitemOpen
  \bibfield  {author} {\bibinfo {author} {\bibfnamefont {R.}~\bibnamefont
  {Albert}}\ and\ \bibinfo {author} {\bibfnamefont {A.-L.}\ \bibnamefont
  {Barab{\'a}si}},\ }\bibfield  {title} {\enquote {\bibinfo {title}
  {Statistical mechanics of complex networks},}\ }\href@noop {} {\bibfield
  {journal} {\bibinfo  {journal} {Reviews of modern physics}\ }\textbf
  {\bibinfo {volume} {74}},\ \bibinfo {pages} {47} (\bibinfo {year}
  {2002})}\BibitemShut {NoStop}%
\bibitem [{\citenamefont {Costa}\ \emph {et~al.}(2007)\citenamefont {Costa},
  \citenamefont {Rodrigues}, \citenamefont {Travieso},\ and\ \citenamefont
  {Villas~Boas}}]{SurveyOfMeasurements}%
  \BibitemOpen
  \bibfield  {author} {\bibinfo {author} {\bibfnamefont {L.~d.~F.}\
  \bibnamefont {Costa}}, \bibinfo {author} {\bibfnamefont {F.~A.}\ \bibnamefont
  {Rodrigues}}, \bibinfo {author} {\bibfnamefont {G.}~\bibnamefont {Travieso}},
  \ and\ \bibinfo {author} {\bibfnamefont {P.}~\bibnamefont {Villas~Boas}},\
  }\bibfield  {title} {\enquote {\bibinfo {title} {Characterization of complex
  networks: A survey of measurements},}\ }\href@noop {} {\bibfield  {journal}
  {\bibinfo  {journal} {Advances in Physics}\ }\textbf {\bibinfo {volume}
  {56}},\ \bibinfo {pages} {167--242} (\bibinfo {year} {2007})}\BibitemShut
  {NoStop}%
\bibitem [{\citenamefont {Christensen}\ and\ \citenamefont
  {Albert}(2007)}]{GraphConcepts}%
  \BibitemOpen
  \bibfield  {author} {\bibinfo {author} {\bibfnamefont {C.}~\bibnamefont
  {Christensen}}\ and\ \bibinfo {author} {\bibfnamefont {R.}~\bibnamefont
  {Albert}},\ }\bibfield  {title} {\enquote {\bibinfo {title} {Using graph
  concepts to understand the organization of complex systems},}\ }\href@noop {}
  {\bibfield  {journal} {\bibinfo  {journal} {International Journal of
  Bifurcation and Chaos}\ }\textbf {\bibinfo {volume} {17}},\ \bibinfo {pages}
  {2201--2214} (\bibinfo {year} {2007})}\BibitemShut {NoStop}%
\bibitem [{\citenamefont {Boccaletti}\ \emph {et~al.}(2006)\citenamefont
  {Boccaletti}, \citenamefont {Latora}, \citenamefont {Moreno}, \citenamefont
  {Chavez},\ and\ \citenamefont {Hwang}}]{StructDynam}%
  \BibitemOpen
  \bibfield  {author} {\bibinfo {author} {\bibfnamefont {S.}~\bibnamefont
  {Boccaletti}}, \bibinfo {author} {\bibfnamefont {V.}~\bibnamefont {Latora}},
  \bibinfo {author} {\bibfnamefont {Y.}~\bibnamefont {Moreno}}, \bibinfo
  {author} {\bibfnamefont {M.}~\bibnamefont {Chavez}}, \ and\ \bibinfo {author}
  {\bibfnamefont {D.-U.}\ \bibnamefont {Hwang}},\ }\bibfield  {title} {\enquote
  {\bibinfo {title} {Complex networks: Structure and dynamics},}\ }\href@noop
  {} {\bibfield  {journal} {\bibinfo  {journal} {Physics reports}\ }\textbf
  {\bibinfo {volume} {424}},\ \bibinfo {pages} {175--308} (\bibinfo {year}
  {2006})}\BibitemShut {NoStop}%
\bibitem [{\citenamefont {Hlinka}, \citenamefont {Hartman},\ and\ \citenamefont
  {Palu{\v{s}}}(2012)}]{DynamicalSys}%
  \BibitemOpen
  \bibfield  {author} {\bibinfo {author} {\bibfnamefont {J.}~\bibnamefont
  {Hlinka}}, \bibinfo {author} {\bibfnamefont {D.}~\bibnamefont {Hartman}}, \
  and\ \bibinfo {author} {\bibfnamefont {M.}~\bibnamefont {Palu{\v{s}}}},\
  }\bibfield  {title} {\enquote {\bibinfo {title} {Small-world topology of
  functional connectivity in randomly connected dynamical systems},}\
  }\href@noop {} {\bibfield  {journal} {\bibinfo  {journal} {Chaos: An
  Interdisciplinary Journal of Nonlinear Science}\ }\textbf {\bibinfo {volume}
  {22}},\ \bibinfo {pages} {033107--033107} (\bibinfo {year}
  {2012})}\BibitemShut {NoStop}%
\bibitem [{\citenamefont {Yazdani}\ and\ \citenamefont
  {Jeffrey}(2011)}]{WaterDistribution}%
  \BibitemOpen
  \bibfield  {author} {\bibinfo {author} {\bibfnamefont {A.}~\bibnamefont
  {Yazdani}}\ and\ \bibinfo {author} {\bibfnamefont {P.}~\bibnamefont
  {Jeffrey}},\ }\bibfield  {title} {\enquote {\bibinfo {title} {Complex network
  analysis of water distribution systems},}\ }\href@noop {} {\bibfield
  {journal} {\bibinfo  {journal} {Chaos: An Interdisciplinary Journal of
  Nonlinear Science}\ }\textbf {\bibinfo {volume} {21}},\ \bibinfo {pages}
  {016111--016111} (\bibinfo {year} {2011})}\BibitemShut {NoStop}%
\bibitem [{\citenamefont {Cano}\ \emph {et~al.}(2006)\citenamefont {Cano},
  \citenamefont {Celma}, \citenamefont {Koppenberger},\ and\ \citenamefont
  {Buld{\'u}}}]{MusicRecom}%
  \BibitemOpen
  \bibfield  {author} {\bibinfo {author} {\bibfnamefont {P.}~\bibnamefont
  {Cano}}, \bibinfo {author} {\bibfnamefont {O.}~\bibnamefont {Celma}},
  \bibinfo {author} {\bibfnamefont {M.}~\bibnamefont {Koppenberger}}, \ and\
  \bibinfo {author} {\bibfnamefont {J.~M.}\ \bibnamefont {Buld{\'u}}},\
  }\bibfield  {title} {\enquote {\bibinfo {title} {Topology of music
  recommendation networks},}\ }\href@noop {} {\bibfield  {journal} {\bibinfo
  {journal} {Chaos: An Interdisciplinary Journal of Nonlinear Science}\
  }\textbf {\bibinfo {volume} {16}},\ \bibinfo {pages} {013107--013107}
  (\bibinfo {year} {2006})}\BibitemShut {NoStop}%
\bibitem [{\citenamefont {Leskovec}\ \emph {et~al.}(2010)\citenamefont
  {Leskovec}, \citenamefont {Chakrabarti}, \citenamefont {Kleinberg},
  \citenamefont {Faloutsos},\ and\ \citenamefont {Ghahramani}}]{Kronecker}%
  \BibitemOpen
  \bibfield  {author} {\bibinfo {author} {\bibfnamefont {J.}~\bibnamefont
  {Leskovec}}, \bibinfo {author} {\bibfnamefont {D.}~\bibnamefont
  {Chakrabarti}}, \bibinfo {author} {\bibfnamefont {J.}~\bibnamefont
  {Kleinberg}}, \bibinfo {author} {\bibfnamefont {C.}~\bibnamefont
  {Faloutsos}}, \ and\ \bibinfo {author} {\bibfnamefont {Z.}~\bibnamefont
  {Ghahramani}},\ }\bibfield  {title} {\enquote {\bibinfo {title} {Kronecker
  graphs: An approach to modeling networks},}\ }\href@noop {} {\bibfield
  {journal} {\bibinfo  {journal} {The Journal of Machine Learning Research}\
  }\textbf {\bibinfo {volume} {11}},\ \bibinfo {pages} {985--1042} (\bibinfo
  {year} {2010})}\BibitemShut {NoStop}%
\bibitem [{\citenamefont {Akoglu}\ and\ \citenamefont {Faloutsos}(2009)}]{RTG}%
  \BibitemOpen
  \bibfield  {author} {\bibinfo {author} {\bibfnamefont {L.}~\bibnamefont
  {Akoglu}}\ and\ \bibinfo {author} {\bibfnamefont {C.}~\bibnamefont
  {Faloutsos}},\ }\bibfield  {title} {\enquote {\bibinfo {title} {Rtg: a
  recursive realistic graph generator using random typing},}\ }\href@noop {}
  {\bibfield  {journal} {\bibinfo  {journal} {Data Mining and Knowledge
  Discovery}\ }\textbf {\bibinfo {volume} {19}},\ \bibinfo {pages} {194--209}
  (\bibinfo {year} {2009})}\BibitemShut {NoStop}%
\bibitem [{\citenamefont {Janssen}, \citenamefont {Hurshman},\ and\
  \citenamefont {Kalyaniwalla}(2012)}]{ModelSelection}%
  \BibitemOpen
  \bibfield  {author} {\bibinfo {author} {\bibfnamefont {J.}~\bibnamefont
  {Janssen}}, \bibinfo {author} {\bibfnamefont {M.}~\bibnamefont {Hurshman}}, \
  and\ \bibinfo {author} {\bibfnamefont {N.}~\bibnamefont {Kalyaniwalla}},\
  }\bibfield  {title} {\enquote {\bibinfo {title} {Model selection for social
  networks using graphlets},}\ }\href@noop {} {\bibfield  {journal} {\bibinfo
  {journal} {Internet Mathematics}\ }\textbf {\bibinfo {volume} {8}},\ \bibinfo
  {pages} {338--363} (\bibinfo {year} {2012})}\BibitemShut {NoStop}%
\bibitem [{\citenamefont {Airoldi}, \citenamefont {Bai},\ and\ \citenamefont
  {Carley}(2011)}]{NetSamplingClassification}%
  \BibitemOpen
  \bibfield  {author} {\bibinfo {author} {\bibfnamefont {E.~M.}\ \bibnamefont
  {Airoldi}}, \bibinfo {author} {\bibfnamefont {X.}~\bibnamefont {Bai}}, \ and\
  \bibinfo {author} {\bibfnamefont {K.~M.}\ \bibnamefont {Carley}},\ }\bibfield
   {title} {\enquote {\bibinfo {title} {Network sampling and classification: An
  investigation of network model representations},}\ }\href@noop {} {\bibfield
  {journal} {\bibinfo  {journal} {Decision Support Systems}\ }\textbf {\bibinfo
  {volume} {51}},\ \bibinfo {pages} {506--518} (\bibinfo {year}
  {2011})}\BibitemShut {NoStop}%
\bibitem [{\citenamefont {Middendorf}, \citenamefont {Ziv},\ and\ \citenamefont
  {Wiggins}(2005)}]{Drosophila}%
  \BibitemOpen
  \bibfield  {author} {\bibinfo {author} {\bibfnamefont {M.}~\bibnamefont
  {Middendorf}}, \bibinfo {author} {\bibfnamefont {E.}~\bibnamefont {Ziv}}, \
  and\ \bibinfo {author} {\bibfnamefont {C.~H.}\ \bibnamefont {Wiggins}},\
  }\bibfield  {title} {\enquote {\bibinfo {title} {Inferring network
  mechanisms: the drosophila melanogaster protein interaction network},}\
  }\href@noop {} {\bibfield  {journal} {\bibinfo  {journal} {Proceedings of the
  National Academy of Sciences of the United States of America}\ }\textbf
  {\bibinfo {volume} {102}},\ \bibinfo {pages} {3192--3197} (\bibinfo {year}
  {2005})}\BibitemShut {NoStop}%
\bibitem [{\citenamefont {Milo}\ \emph {et~al.}(2004)\citenamefont {Milo},
  \citenamefont {Itzkovitz}, \citenamefont {Kashtan}, \citenamefont {Levitt},
  \citenamefont {Shen-Orr}, \citenamefont {Ayzenshtat}, \citenamefont
  {Sheffer},\ and\ \citenamefont {Alon}}]{Superfamilies}%
  \BibitemOpen
  \bibfield  {author} {\bibinfo {author} {\bibfnamefont {R.}~\bibnamefont
  {Milo}}, \bibinfo {author} {\bibfnamefont {S.}~\bibnamefont {Itzkovitz}},
  \bibinfo {author} {\bibfnamefont {N.}~\bibnamefont {Kashtan}}, \bibinfo
  {author} {\bibfnamefont {R.}~\bibnamefont {Levitt}}, \bibinfo {author}
  {\bibfnamefont {S.}~\bibnamefont {Shen-Orr}}, \bibinfo {author}
  {\bibfnamefont {I.}~\bibnamefont {Ayzenshtat}}, \bibinfo {author}
  {\bibfnamefont {M.}~\bibnamefont {Sheffer}}, \ and\ \bibinfo {author}
  {\bibfnamefont {U.}~\bibnamefont {Alon}},\ }\bibfield  {title} {\enquote
  {\bibinfo {title} {Superfamilies of evolved and designed networks},}\
  }\href@noop {} {\bibfield  {journal} {\bibinfo  {journal} {Science}\ }\textbf
  {\bibinfo {volume} {303}},\ \bibinfo {pages} {1538--1542} (\bibinfo {year}
  {2004})}\BibitemShut {NoStop}%
\bibitem [{\citenamefont {Milo}\ \emph {et~al.}(2002)\citenamefont {Milo},
  \citenamefont {Shen-Orr}, \citenamefont {Itzkovitz}, \citenamefont {Kashtan},
  \citenamefont {Chklovskii},\ and\ \citenamefont {Alon}}]{Motifs}%
  \BibitemOpen
  \bibfield  {author} {\bibinfo {author} {\bibfnamefont {R.}~\bibnamefont
  {Milo}}, \bibinfo {author} {\bibfnamefont {S.}~\bibnamefont {Shen-Orr}},
  \bibinfo {author} {\bibfnamefont {S.}~\bibnamefont {Itzkovitz}}, \bibinfo
  {author} {\bibfnamefont {N.}~\bibnamefont {Kashtan}}, \bibinfo {author}
  {\bibfnamefont {D.}~\bibnamefont {Chklovskii}}, \ and\ \bibinfo {author}
  {\bibfnamefont {U.}~\bibnamefont {Alon}},\ }\bibfield  {title} {\enquote
  {\bibinfo {title} {Network motifs: simple building blocks of complex
  networks},}\ }\href@noop {} {\bibfield  {journal} {\bibinfo  {journal}
  {Science Signaling}\ }\textbf {\bibinfo {volume} {298}},\ \bibinfo {pages}
  {824} (\bibinfo {year} {2002})}\BibitemShut {NoStop}%
\bibitem [{\citenamefont {Bordino}\ \emph {et~al.}(2008)\citenamefont
  {Bordino}, \citenamefont {Donato}, \citenamefont {Gionis},\ and\
  \citenamefont {Leonardi}}]{SubgraphCounting}%
  \BibitemOpen
  \bibfield  {author} {\bibinfo {author} {\bibfnamefont {I.}~\bibnamefont
  {Bordino}}, \bibinfo {author} {\bibfnamefont {D.}~\bibnamefont {Donato}},
  \bibinfo {author} {\bibfnamefont {A.}~\bibnamefont {Gionis}}, \ and\ \bibinfo
  {author} {\bibfnamefont {S.}~\bibnamefont {Leonardi}},\ }\bibfield  {title}
  {\enquote {\bibinfo {title} {Mining large networks with subgraph counting},}\
  }in\ \href@noop {} {\emph {\bibinfo {booktitle} {Data Mining, 2008. ICDM'08.
  Eighth IEEE International Conference on}}}\ (\bibinfo {organization} {IEEE},\
  \bibinfo {year} {2008})\ pp.\ \bibinfo {pages} {737--742}\BibitemShut
  {NoStop}%
\bibitem [{\citenamefont {Grochow}\ and\ \citenamefont
  {Kellis}(2007)}]{NetMotifDiscovery}%
  \BibitemOpen
  \bibfield  {author} {\bibinfo {author} {\bibfnamefont {J.~A.}\ \bibnamefont
  {Grochow}}\ and\ \bibinfo {author} {\bibfnamefont {M.}~\bibnamefont
  {Kellis}},\ }\bibfield  {title} {\enquote {\bibinfo {title} {Network motif
  discovery using subgraph enumeration and symmetry-breaking},}\ }in\
  \href@noop {} {\emph {\bibinfo {booktitle} {Research in Computational
  Molecular Biology}}}\ (\bibinfo {organization} {Springer},\ \bibinfo {year}
  {2007})\ pp.\ \bibinfo {pages} {92--106}\BibitemShut {NoStop}%
\bibitem [{\citenamefont {Cui}, \citenamefont {Kumara},\ and\ \citenamefont
  {Albert}(2010)}]{EngineeringView}%
  \BibitemOpen
  \bibfield  {author} {\bibinfo {author} {\bibfnamefont {L.}~\bibnamefont
  {Cui}}, \bibinfo {author} {\bibfnamefont {S.}~\bibnamefont {Kumara}}, \ and\
  \bibinfo {author} {\bibfnamefont {R.}~\bibnamefont {Albert}},\ }\bibfield
  {title} {\enquote {\bibinfo {title} {Complex networks: An engineering
  view},}\ }\href@noop {} {\bibfield  {journal} {\bibinfo  {journal} {Circuits
  and Systems Magazine, IEEE}\ }\textbf {\bibinfo {volume} {10}},\ \bibinfo
  {pages} {10--25} (\bibinfo {year} {2010})}\BibitemShut {NoStop}%
\bibitem [{\citenamefont {Pomerance}\ \emph {et~al.}(2009)\citenamefont
  {Pomerance}, \citenamefont {Ott}, \citenamefont {Girvan},\ and\ \citenamefont
  {Losert}}]{EffectOfNetTopology}%
  \BibitemOpen
  \bibfield  {author} {\bibinfo {author} {\bibfnamefont {A.}~\bibnamefont
  {Pomerance}}, \bibinfo {author} {\bibfnamefont {E.}~\bibnamefont {Ott}},
  \bibinfo {author} {\bibfnamefont {M.}~\bibnamefont {Girvan}}, \ and\ \bibinfo
  {author} {\bibfnamefont {W.}~\bibnamefont {Losert}},\ }\bibfield  {title}
  {\enquote {\bibinfo {title} {The effect of network topology on the stability
  of discrete state models of genetic control},}\ }\href@noop {} {\bibfield
  {journal} {\bibinfo  {journal} {Proceedings of the National Academy of
  Sciences}\ }\textbf {\bibinfo {volume} {106}},\ \bibinfo {pages} {8209--8214}
  (\bibinfo {year} {2009})}\BibitemShut {NoStop}%
\bibitem [{\citenamefont {Lameu}\ \emph {et~al.}(2012)\citenamefont {Lameu},
  \citenamefont {Batista}, \citenamefont {Batista}, \citenamefont {Iarosz},
  \citenamefont {Viana}, \citenamefont {Lopes},\ and\ \citenamefont
  {Kurths}}]{RichClub}%
  \BibitemOpen
  \bibfield  {author} {\bibinfo {author} {\bibfnamefont {E.}~\bibnamefont
  {Lameu}}, \bibinfo {author} {\bibfnamefont {C.}~\bibnamefont {Batista}},
  \bibinfo {author} {\bibfnamefont {A.}~\bibnamefont {Batista}}, \bibinfo
  {author} {\bibfnamefont {K.}~\bibnamefont {Iarosz}}, \bibinfo {author}
  {\bibfnamefont {R.}~\bibnamefont {Viana}}, \bibinfo {author} {\bibfnamefont
  {S.}~\bibnamefont {Lopes}}, \ and\ \bibinfo {author} {\bibfnamefont
  {J.}~\bibnamefont {Kurths}},\ }\bibfield  {title} {\enquote {\bibinfo {title}
  {Suppression of bursting synchronization in clustered scale-free (rich-club)
  neuronal networks},}\ }\href@noop {} {\bibfield  {journal} {\bibinfo
  {journal} {Chaos: An Interdisciplinary Journal of Nonlinear Science}\
  }\textbf {\bibinfo {volume} {22}},\ \bibinfo {pages} {043149} (\bibinfo
  {year} {2012})}\BibitemShut {NoStop}%
\bibitem [{\citenamefont {Leskovec}, \citenamefont {Kleinberg},\ and\
  \citenamefont {Faloutsos}(2005)}]{GraphsOverTime}%
  \BibitemOpen
  \bibfield  {author} {\bibinfo {author} {\bibfnamefont {J.}~\bibnamefont
  {Leskovec}}, \bibinfo {author} {\bibfnamefont {J.}~\bibnamefont {Kleinberg}},
  \ and\ \bibinfo {author} {\bibfnamefont {C.}~\bibnamefont {Faloutsos}},\
  }\bibfield  {title} {\enquote {\bibinfo {title} {Graphs over time:
  densification laws, shrinking diameters and possible explanations},}\ }in\
  \href@noop {} {\emph {\bibinfo {booktitle} {Proceedings of the eleventh ACM
  SIGKDD international conference on Knowledge discovery in data mining}}}\
  (\bibinfo {organization} {ACM},\ \bibinfo {year} {2005})\ pp.\ \bibinfo
  {pages} {177--187}\BibitemShut {NoStop}%
\bibitem [{\citenamefont {Kleinberg}\ \emph {et~al.}(1999)\citenamefont
  {Kleinberg}, \citenamefont {Kumar}, \citenamefont {Raghavan}, \citenamefont
  {Rajagopalan},\ and\ \citenamefont {Tomkins}}]{WebAsGraph}%
  \BibitemOpen
  \bibfield  {author} {\bibinfo {author} {\bibfnamefont {J.~M.}\ \bibnamefont
  {Kleinberg}}, \bibinfo {author} {\bibfnamefont {R.}~\bibnamefont {Kumar}},
  \bibinfo {author} {\bibfnamefont {P.}~\bibnamefont {Raghavan}}, \bibinfo
  {author} {\bibfnamefont {S.}~\bibnamefont {Rajagopalan}}, \ and\ \bibinfo
  {author} {\bibfnamefont {A.~S.}\ \bibnamefont {Tomkins}},\ }\bibfield
  {title} {\enquote {\bibinfo {title} {The web as a graph: Measurements,
  models, and methods},}\ }in\ \href@noop {} {\emph {\bibinfo {booktitle}
  {Computing and combinatorics}}}\ (\bibinfo  {publisher} {Springer},\ \bibinfo
  {year} {1999})\ pp.\ \bibinfo {pages} {1--17}\BibitemShut {NoStop}%
\bibitem [{\citenamefont {Barab{\'a}si}\ and\ \citenamefont
  {Albert}(1999)}]{EmergenceOfScaling}%
  \BibitemOpen
  \bibfield  {author} {\bibinfo {author} {\bibfnamefont {A.-L.}\ \bibnamefont
  {Barab{\'a}si}}\ and\ \bibinfo {author} {\bibfnamefont {R.}~\bibnamefont
  {Albert}},\ }\bibfield  {title} {\enquote {\bibinfo {title} {Emergence of
  scaling in random networks},}\ }\href@noop {} {\bibfield  {journal} {\bibinfo
   {journal} {science}\ }\textbf {\bibinfo {volume} {286}},\ \bibinfo {pages}
  {509--512} (\bibinfo {year} {1999})}\BibitemShut {NoStop}%
\bibitem [{\citenamefont {Watts}\ and\ \citenamefont
  {Strogatz}(1998)}]{SmallWorld}%
  \BibitemOpen
  \bibfield  {author} {\bibinfo {author} {\bibfnamefont {D.~J.}\ \bibnamefont
  {Watts}}\ and\ \bibinfo {author} {\bibfnamefont {S.~H.}\ \bibnamefont
  {Strogatz}},\ }\bibfield  {title} {\enquote {\bibinfo {title} {Collective
  dynamics of `small-world'networks},}\ }\href@noop {} {\bibfield  {journal}
  {\bibinfo  {journal} {nature}\ }\textbf {\bibinfo {volume} {393}},\ \bibinfo
  {pages} {440--442} (\bibinfo {year} {1998})}\BibitemShut {NoStop}%
\bibitem [{\citenamefont {Erd{\"o}s}\ and\ \citenamefont
  {R{\'e}nyi}(1959)}]{ER}%
  \BibitemOpen
  \bibfield  {author} {\bibinfo {author} {\bibfnamefont {P.}~\bibnamefont
  {Erd{\"o}s}}\ and\ \bibinfo {author} {\bibfnamefont {A.}~\bibnamefont
  {R{\'e}nyi}},\ }\bibfield  {title} {\enquote {\bibinfo {title} {On the
  central limit theorem for samples from a finite population},}\ }\href@noop {}
  {\bibfield  {journal} {\bibinfo  {journal} {Publ. Math. Inst. Hungar. Acad.
  Sci}\ }\textbf {\bibinfo {volume} {4}},\ \bibinfo {pages} {49--61} (\bibinfo
  {year} {1959})}\BibitemShut {NoStop}%
\bibitem [{\citenamefont {Volchenkov}\ and\ \citenamefont
  {Blanchard}(2002)}]{RP}%
  \BibitemOpen
  \bibfield  {author} {\bibinfo {author} {\bibfnamefont {D.}~\bibnamefont
  {Volchenkov}}\ and\ \bibinfo {author} {\bibfnamefont {P.}~\bibnamefont
  {Blanchard}},\ }\bibfield  {title} {\enquote {\bibinfo {title} {An algorithm
  generating random graphs with power law degree distributions},}\ }\href@noop
  {} {\bibfield  {journal} {\bibinfo  {journal} {Physica A: Statistical
  Mechanics and its Applications}\ }\textbf {\bibinfo {volume} {315}},\
  \bibinfo {pages} {677--690} (\bibinfo {year} {2002})}\BibitemShut {NoStop}%
\bibitem [{\citenamefont {Penrose}(2003)}]{RandomGeoGraphs}%
  \BibitemOpen
  \bibfield  {author} {\bibinfo {author} {\bibfnamefont {M.}~\bibnamefont
  {Penrose}},\ }\href@noop {} {\emph {\bibinfo {title} {Random geometric
  graphs}}},\ Vol.~\bibinfo {volume} {5}\ (\bibinfo  {publisher} {Oxford
  University Press Oxford},\ \bibinfo {year} {2003})\BibitemShut {NoStop}%
\bibitem [{\citenamefont {Aiello}\ \emph {et~al.}(2008)\citenamefont {Aiello},
  \citenamefont {Bonato}, \citenamefont {Cooper}, \citenamefont {Janssen},\
  and\ \citenamefont {Pra{\l}at}}]{SpatialWebGraph}%
  \BibitemOpen
  \bibfield  {author} {\bibinfo {author} {\bibfnamefont {W.}~\bibnamefont
  {Aiello}}, \bibinfo {author} {\bibfnamefont {A.}~\bibnamefont {Bonato}},
  \bibinfo {author} {\bibfnamefont {C.}~\bibnamefont {Cooper}}, \bibinfo
  {author} {\bibfnamefont {J.}~\bibnamefont {Janssen}}, \ and\ \bibinfo
  {author} {\bibfnamefont {P.}~\bibnamefont {Pra{\l}at}},\ }\bibfield  {title}
  {\enquote {\bibinfo {title} {A spatial web graph model with local influence
  regions},}\ }\href@noop {} {\bibfield  {journal} {\bibinfo  {journal}
  {Internet Mathematics}\ }\textbf {\bibinfo {volume} {5}},\ \bibinfo {pages}
  {175--196} (\bibinfo {year} {2008})}\BibitemShut {NoStop}%
\bibitem [{\citenamefont {Callaway}\ \emph {et~al.}(2001)\citenamefont
  {Callaway}, \citenamefont {Hopcroft}, \citenamefont {Kleinberg},
  \citenamefont {Newman},\ and\ \citenamefont {Strogatz}}]{RandomlyGrown}%
  \BibitemOpen
  \bibfield  {author} {\bibinfo {author} {\bibfnamefont {D.~S.}\ \bibnamefont
  {Callaway}}, \bibinfo {author} {\bibfnamefont {J.~E.}\ \bibnamefont
  {Hopcroft}}, \bibinfo {author} {\bibfnamefont {J.~M.}\ \bibnamefont
  {Kleinberg}}, \bibinfo {author} {\bibfnamefont {M.~E.}\ \bibnamefont
  {Newman}}, \ and\ \bibinfo {author} {\bibfnamefont {S.~H.}\ \bibnamefont
  {Strogatz}},\ }\bibfield  {title} {\enquote {\bibinfo {title} {Are randomly
  grown graphs really random?}}\ }\href@noop {} {\bibfield  {journal} {\bibinfo
   {journal} {Physical Review E}\ }\textbf {\bibinfo {volume} {64}},\ \bibinfo
  {pages} {041902} (\bibinfo {year} {2001})}\BibitemShut {NoStop}%
\bibitem [{\citenamefont {Sol{\'e}}\ \emph {et~al.}(2002)\citenamefont
  {Sol{\'e}}, \citenamefont {Pastor-Satorras}, \citenamefont {Smith},\ and\
  \citenamefont {Kepler}}]{ProteomeEvolution}%
  \BibitemOpen
  \bibfield  {author} {\bibinfo {author} {\bibfnamefont {R.~V.}\ \bibnamefont
  {Sol{\'e}}}, \bibinfo {author} {\bibfnamefont {R.}~\bibnamefont
  {Pastor-Satorras}}, \bibinfo {author} {\bibfnamefont {E.}~\bibnamefont
  {Smith}}, \ and\ \bibinfo {author} {\bibfnamefont {T.~B.}\ \bibnamefont
  {Kepler}},\ }\bibfield  {title} {\enquote {\bibinfo {title} {A model of
  large-scale proteome evolution},}\ }\href@noop {} {\bibfield  {journal}
  {\bibinfo  {journal} {Advances in Complex Systems}\ }\textbf {\bibinfo
  {volume} {5}},\ \bibinfo {pages} {43--54} (\bibinfo {year}
  {2002})}\BibitemShut {NoStop}%
\bibitem [{\citenamefont {Klemm}\ and\ \citenamefont
  {Eguiluz}(2002)}]{HighlyClustered}%
  \BibitemOpen
  \bibfield  {author} {\bibinfo {author} {\bibfnamefont {K.}~\bibnamefont
  {Klemm}}\ and\ \bibinfo {author} {\bibfnamefont {V.~M.}\ \bibnamefont
  {Eguiluz}},\ }\bibfield  {title} {\enquote {\bibinfo {title} {Highly
  clustered scale-free networks},}\ }\href@noop {} {\bibfield  {journal}
  {\bibinfo  {journal} {Physical Review E}\ }\textbf {\bibinfo {volume} {65}},\
  \bibinfo {pages} {036123} (\bibinfo {year} {2002})}\BibitemShut {NoStop}%
\bibitem [{\citenamefont {Bollob{\'a}s}(2001)}]{RandomGraphs}%
  \BibitemOpen
  \bibfield  {author} {\bibinfo {author} {\bibfnamefont {B.}~\bibnamefont
  {Bollob{\'a}s}},\ }\href@noop {} {\emph {\bibinfo {title} {Random graphs}}},\
  Vol.~\bibinfo {volume} {73}\ (\bibinfo  {publisher} {Cambridge university
  press},\ \bibinfo {year} {2001})\BibitemShut {NoStop}%
\bibitem [{\citenamefont {Borgatti}\ and\ \citenamefont
  {Everett}(2000)}]{CorePeriphery}%
  \BibitemOpen
  \bibfield  {author} {\bibinfo {author} {\bibfnamefont {S.~P.}\ \bibnamefont
  {Borgatti}}\ and\ \bibinfo {author} {\bibfnamefont {M.~G.}\ \bibnamefont
  {Everett}},\ }\bibfield  {title} {\enquote {\bibinfo {title} {Models of
  core/periphery structures},}\ }\href@noop {} {\bibfield  {journal} {\bibinfo
  {journal} {Social networks}\ }\textbf {\bibinfo {volume} {21}},\ \bibinfo
  {pages} {375--395} (\bibinfo {year} {2000})}\BibitemShut {NoStop}%
\bibitem [{\citenamefont {Frantz}\ and\ \citenamefont
  {Carley}(2005)}]{CellularNets}%
  \BibitemOpen
  \bibfield  {author} {\bibinfo {author} {\bibfnamefont {T.~L.}\ \bibnamefont
  {Frantz}}\ and\ \bibinfo {author} {\bibfnamefont {K.~M.}\ \bibnamefont
  {Carley}},\ }\href@noop {} {\enquote {\bibinfo {title} {A formal
  characterization of cellular networks},}\ }\bibinfo {type} {Tech. Rep.}\
  (\bibinfo  {institution} {DTIC Document},\ \bibinfo {year}
  {2005})\BibitemShut {NoStop}%
\bibitem [{\citenamefont {Leskovec}, \citenamefont {Kleinberg},\ and\
  \citenamefont {Faloutsos}(2007)}]{GraphEvolution}%
  \BibitemOpen
  \bibfield  {author} {\bibinfo {author} {\bibfnamefont {J.}~\bibnamefont
  {Leskovec}}, \bibinfo {author} {\bibfnamefont {J.}~\bibnamefont {Kleinberg}},
  \ and\ \bibinfo {author} {\bibfnamefont {C.}~\bibnamefont {Faloutsos}},\
  }\bibfield  {title} {\enquote {\bibinfo {title} {Graph evolution:
  Densification and shrinking diameters},}\ }\href@noop {} {\bibfield
  {journal} {\bibinfo  {journal} {ACM Transactions on Knowledge Discovery from
  Data (TKDD)}\ }\textbf {\bibinfo {volume} {1}},\ \bibinfo {pages} {2}
  (\bibinfo {year} {2007})}\BibitemShut {NoStop}%
\bibitem [{\citenamefont {Holmes}\ \emph {et~al.}(2002)\citenamefont {Holmes},
  \citenamefont {Pfahringer}, \citenamefont {Kirkby}, \citenamefont {Frank},\
  and\ \citenamefont {Hall}}]{MulticlassADT}%
  \BibitemOpen
  \bibfield  {author} {\bibinfo {author} {\bibfnamefont {G.}~\bibnamefont
  {Holmes}}, \bibinfo {author} {\bibfnamefont {B.}~\bibnamefont {Pfahringer}},
  \bibinfo {author} {\bibfnamefont {R.}~\bibnamefont {Kirkby}}, \bibinfo
  {author} {\bibfnamefont {E.}~\bibnamefont {Frank}}, \ and\ \bibinfo {author}
  {\bibfnamefont {M.}~\bibnamefont {Hall}},\ }\bibfield  {title} {\enquote
  {\bibinfo {title} {Multiclass alternating decision trees},}\ }in\ \href@noop
  {} {\emph {\bibinfo {booktitle} {Machine Learning: ECML 2002}}}\ (\bibinfo
  {publisher} {Springer},\ \bibinfo {year} {2002})\ pp.\ \bibinfo {pages}
  {161--172}\BibitemShut {NoStop}%
\bibitem [{\citenamefont {Patro}\ \emph {et~al.}(2012)\citenamefont {Patro},
  \citenamefont {Duggal}, \citenamefont {Sefer}, \citenamefont {Wang},
  \citenamefont {Filippova},\ and\ \citenamefont {Kingsford}}]{MissingModels}%
  \BibitemOpen
  \bibfield  {author} {\bibinfo {author} {\bibfnamefont {R.}~\bibnamefont
  {Patro}}, \bibinfo {author} {\bibfnamefont {G.}~\bibnamefont {Duggal}},
  \bibinfo {author} {\bibfnamefont {E.}~\bibnamefont {Sefer}}, \bibinfo
  {author} {\bibfnamefont {H.}~\bibnamefont {Wang}}, \bibinfo {author}
  {\bibfnamefont {D.}~\bibnamefont {Filippova}}, \ and\ \bibinfo {author}
  {\bibfnamefont {C.}~\bibnamefont {Kingsford}},\ }\bibfield  {title} {\enquote
  {\bibinfo {title} {The missing models: a data-driven approach for learning
  how networks grow},}\ }in\ \href@noop {} {\emph {\bibinfo {booktitle}
  {Proceedings of the 18th ACM SIGKDD international conference on Knowledge
  discovery and data mining}}}\ (\bibinfo {organization} {ACM},\ \bibinfo
  {year} {2012})\ pp.\ \bibinfo {pages} {42--50}\BibitemShut {NoStop}%
\bibitem [{\citenamefont {Zanin}\ \emph {et~al.}(2012)\citenamefont {Zanin},
  \citenamefont {Sousa}, \citenamefont {Papo}, \citenamefont {Bajo},
  \citenamefont {Garc{\'\i}a-Prieto}, \citenamefont {del Pozo}, \citenamefont
  {Menasalvas},\ and\ \citenamefont {Boccaletti}}]{Zanin}%
  \BibitemOpen
  \bibfield  {author} {\bibinfo {author} {\bibfnamefont {M.}~\bibnamefont
  {Zanin}}, \bibinfo {author} {\bibfnamefont {P.}~\bibnamefont {Sousa}},
  \bibinfo {author} {\bibfnamefont {D.}~\bibnamefont {Papo}}, \bibinfo {author}
  {\bibfnamefont {R.}~\bibnamefont {Bajo}}, \bibinfo {author} {\bibfnamefont
  {J.}~\bibnamefont {Garc{\'\i}a-Prieto}}, \bibinfo {author} {\bibfnamefont
  {F.}~\bibnamefont {del Pozo}}, \bibinfo {author} {\bibfnamefont
  {E.}~\bibnamefont {Menasalvas}}, \ and\ \bibinfo {author} {\bibfnamefont
  {S.}~\bibnamefont {Boccaletti}},\ }\bibfield  {title} {\enquote {\bibinfo
  {title} {Optimizing functional network representation of multivariate time
  series},}\ }\href@noop {} {\bibfield  {journal} {\bibinfo  {journal}
  {Scientific reports}\ }\textbf {\bibinfo {volume} {2}} (\bibinfo {year}
  {2012})}\BibitemShut {NoStop}%
\bibitem [{\citenamefont {Barrat}\ and\ \citenamefont
  {Weigt}(2000)}]{TransitivityProp}%
  \BibitemOpen
  \bibfield  {author} {\bibinfo {author} {\bibfnamefont {A.}~\bibnamefont
  {Barrat}}\ and\ \bibinfo {author} {\bibfnamefont {M.}~\bibnamefont {Weigt}},\
  }\bibfield  {title} {\enquote {\bibinfo {title} {On the properties of
  small-world network models},}\ }\href@noop {} {\bibfield  {journal} {\bibinfo
   {journal} {The European Physical Journal B-Condensed Matter and Complex
  Systems}\ }\textbf {\bibinfo {volume} {13}},\ \bibinfo {pages} {547--560}
  (\bibinfo {year} {2000})}\BibitemShut {NoStop}%
\bibitem [{\citenamefont {Bollob{\'a}s}(1981)}]{DiameterProp}%
  \BibitemOpen
  \bibfield  {author} {\bibinfo {author} {\bibfnamefont {B.}~\bibnamefont
  {Bollob{\'a}s}},\ }\bibfield  {title} {\enquote {\bibinfo {title} {The
  diameter of random graphs},}\ }\href@noop {} {\bibfield  {journal} {\bibinfo
  {journal} {Transactions of the American Mathematical Society}\ }\textbf
  {\bibinfo {volume} {267}},\ \bibinfo {pages} {41--52} (\bibinfo {year}
  {1981})}\BibitemShut {NoStop}%
\bibitem [{\citenamefont {Boas}\ \emph {et~al.}(2010)\citenamefont {Boas},
  \citenamefont {Rodrigues}, \citenamefont {Travieso},\ and\ \citenamefont
  {da~F~Costa}}]{Sensitivity}%
  \BibitemOpen
  \bibfield  {author} {\bibinfo {author} {\bibfnamefont {P.~V.}\ \bibnamefont
  {Boas}}, \bibinfo {author} {\bibfnamefont {F.}~\bibnamefont {Rodrigues}},
  \bibinfo {author} {\bibfnamefont {G.}~\bibnamefont {Travieso}}, \ and\
  \bibinfo {author} {\bibfnamefont {L.}~\bibnamefont {da~F~Costa}},\ }\bibfield
   {title} {\enquote {\bibinfo {title} {Sensitivity of complex networks
  measurements},}\ }\href@noop {} {\bibfield  {journal} {\bibinfo  {journal}
  {Journal of Statistical Mechanics: Theory and Experiment}\ }\textbf {\bibinfo
  {volume} {2010}},\ \bibinfo {pages} {P03009} (\bibinfo {year}
  {2010})}\BibitemShut {NoStop}%
\bibitem [{\citenamefont {Tauro}\ \emph {et~al.}(2001)\citenamefont {Tauro},
  \citenamefont {Palmer}, \citenamefont {Siganos},\ and\ \citenamefont
  {Faloutsos}}]{InternetTopology}%
  \BibitemOpen
  \bibfield  {author} {\bibinfo {author} {\bibfnamefont {S.~L.}\ \bibnamefont
  {Tauro}}, \bibinfo {author} {\bibfnamefont {C.}~\bibnamefont {Palmer}},
  \bibinfo {author} {\bibfnamefont {G.}~\bibnamefont {Siganos}}, \ and\
  \bibinfo {author} {\bibfnamefont {M.}~\bibnamefont {Faloutsos}},\ }\bibfield
  {title} {\enquote {\bibinfo {title} {A simple conceptual model for the
  internet topology},}\ }in\ \href@noop {} {\emph {\bibinfo {booktitle} {Global
  Telecommunications Conference, 2001. GLOBECOM'01. IEEE}}},\ Vol.~\bibinfo
  {volume} {3}\ (\bibinfo {organization} {IEEE},\ \bibinfo {year} {2001})\ pp.\
  \bibinfo {pages} {1667--1671}\BibitemShut {NoStop}%
\bibitem [{\citenamefont {G{\'o}mez}, \citenamefont {Kaltenbrunner},\ and\
  \citenamefont {L{\'o}pez}(2008)}]{Slashdot}%
  \BibitemOpen
  \bibfield  {author} {\bibinfo {author} {\bibfnamefont {V.}~\bibnamefont
  {G{\'o}mez}}, \bibinfo {author} {\bibfnamefont {A.}~\bibnamefont
  {Kaltenbrunner}}, \ and\ \bibinfo {author} {\bibfnamefont {V.}~\bibnamefont
  {L{\'o}pez}},\ }\bibfield  {title} {\enquote {\bibinfo {title} {Statistical
  analysis of the social network and discussion threads in slashdot},}\ }in\
  \href@noop {} {\emph {\bibinfo {booktitle} {Proceedings of the 17th
  international conference on World Wide Web}}}\ (\bibinfo {organization}
  {ACM},\ \bibinfo {year} {2008})\ pp.\ \bibinfo {pages} {645--654}\BibitemShut
  {NoStop}%
\bibitem [{\citenamefont {Gong}\ \emph {et~al.}(2012)\citenamefont {Gong},
  \citenamefont {Xu}, \citenamefont {Huang}, \citenamefont {Mittal},
  \citenamefont {Stefanov}, \citenamefont {Sekar},\ and\ \citenamefont
  {Song}}]{GooglePlus}%
  \BibitemOpen
  \bibfield  {author} {\bibinfo {author} {\bibfnamefont {N.~Z.}\ \bibnamefont
  {Gong}}, \bibinfo {author} {\bibfnamefont {W.}~\bibnamefont {Xu}}, \bibinfo
  {author} {\bibfnamefont {L.}~\bibnamefont {Huang}}, \bibinfo {author}
  {\bibfnamefont {P.}~\bibnamefont {Mittal}}, \bibinfo {author} {\bibfnamefont
  {E.}~\bibnamefont {Stefanov}}, \bibinfo {author} {\bibfnamefont
  {V.}~\bibnamefont {Sekar}}, \ and\ \bibinfo {author} {\bibfnamefont
  {D.}~\bibnamefont {Song}},\ }\bibfield  {title} {\enquote {\bibinfo {title}
  {Evolution of social-attribute networks: measurements, modeling, and
  implications using google+},}\ }in\ \href@noop {} {\emph {\bibinfo
  {booktitle} {Proceedings of the 2012 ACM conference on Internet measurement
  conference}}}\ (\bibinfo {organization} {ACM},\ \bibinfo {year} {2012})\ pp.\
  \bibinfo {pages} {131--144}\BibitemShut {NoStop}%
\bibitem [{\citenamefont {Kwak}\ \emph {et~al.}(2010)\citenamefont {Kwak},
  \citenamefont {Lee}, \citenamefont {Park},\ and\ \citenamefont
  {Moon}}]{WhatIsTwitter}%
  \BibitemOpen
  \bibfield  {author} {\bibinfo {author} {\bibfnamefont {H.}~\bibnamefont
  {Kwak}}, \bibinfo {author} {\bibfnamefont {C.}~\bibnamefont {Lee}}, \bibinfo
  {author} {\bibfnamefont {H.}~\bibnamefont {Park}}, \ and\ \bibinfo {author}
  {\bibfnamefont {S.}~\bibnamefont {Moon}},\ }\bibfield  {title} {\enquote
  {\bibinfo {title} {What is twitter, a social network or a news media?}}\ }in\
  \href@noop {} {\emph {\bibinfo {booktitle} {Proceedings of the 19th
  international conference on World wide web}}}\ (\bibinfo {organization}
  {ACM},\ \bibinfo {year} {2010})\ pp.\ \bibinfo {pages} {591--600}\BibitemShut
  {NoStop}%
\bibitem [{\citenamefont {Quinlan}(1993)}]{QC4.5}%
  \BibitemOpen
  \bibfield  {author} {\bibinfo {author} {\bibfnamefont {J.~R.}\ \bibnamefont
  {Quinlan}},\ }\href@noop {} {\emph {\bibinfo {title} {C4. 5: programs for
  machine learning}}},\ Vol.~\bibinfo {volume} {1}\ (\bibinfo  {publisher}
  {Morgan kaufmann},\ \bibinfo {year} {1993})\BibitemShut {NoStop}%
\bibitem [{\citenamefont {Friedman}, \citenamefont {Geiger},\ and\
  \citenamefont {Goldszmidt}(1997)}]{BayesianRef}%
  \BibitemOpen
  \bibfield  {author} {\bibinfo {author} {\bibfnamefont {N.}~\bibnamefont
  {Friedman}}, \bibinfo {author} {\bibfnamefont {D.}~\bibnamefont {Geiger}}, \
  and\ \bibinfo {author} {\bibfnamefont {M.}~\bibnamefont {Goldszmidt}},\
  }\bibfield  {title} {\enquote {\bibinfo {title} {Bayesian network
  classifiers},}\ }\href@noop {} {\bibfield  {journal} {\bibinfo  {journal}
  {Machine learning}\ }\textbf {\bibinfo {volume} {29}},\ \bibinfo {pages}
  {131--163} (\bibinfo {year} {1997})}\BibitemShut {NoStop}%
\bibitem [{\citenamefont {Platt}(1999)}]{SMORef}%
  \BibitemOpen
  \bibfield  {author} {\bibinfo {author} {\bibfnamefont {J.~C.}\ \bibnamefont
  {Platt}},\ }\bibfield  {title} {\enquote {\bibinfo {title} {12 fast training
  of support vector machines using sequential minimal optimization},}\
  }\href@noop {} {\  (\bibinfo {year} {1999})}\BibitemShut {NoStop}%
\bibitem [{\citenamefont {Freeman}\ and\ \citenamefont
  {Skapura}(1991)}]{NeuralNetRef}%
  \BibitemOpen
  \bibfield  {author} {\bibinfo {author} {\bibfnamefont {J.~A.}\ \bibnamefont
  {Freeman}}\ and\ \bibinfo {author} {\bibfnamefont {D.~M.}\ \bibnamefont
  {Skapura}},\ }\bibfield  {title} {\enquote {\bibinfo {title} {Neural
  networks: Algorithms, applications, and programming techniques (computation
  and neural systems series)},}\ }\href@noop {} {\bibfield  {journal} {\bibinfo
   {journal} {Neural networks: algorithms, applications and programming
  techniques (Computation and Neural Systems Series)}\ } (\bibinfo {year}
  {1991})}\BibitemShut {NoStop}%
\bibitem [{\citenamefont {de~Silva}\ and\ \citenamefont
  {Stumpf}(2005)}]{ModelsInBiology}%
  \BibitemOpen
  \bibfield  {author} {\bibinfo {author} {\bibfnamefont {E.}~\bibnamefont
  {de~Silva}}\ and\ \bibinfo {author} {\bibfnamefont {M.~P.}\ \bibnamefont
  {Stumpf}},\ }\bibfield  {title} {\enquote {\bibinfo {title} {Complex networks
  and simple models in biology},}\ }\href@noop {} {\bibfield  {journal}
  {\bibinfo  {journal} {Journal of the Royal Society Interface}\ }\textbf
  {\bibinfo {volume} {2}},\ \bibinfo {pages} {419--430} (\bibinfo {year}
  {2005})}\BibitemShut {NoStop}%
\bibitem [{\citenamefont {Aittokallio}\ and\ \citenamefont
  {Schwikowski}(2006)}]{CellBiology}%
  \BibitemOpen
  \bibfield  {author} {\bibinfo {author} {\bibfnamefont {T.}~\bibnamefont
  {Aittokallio}}\ and\ \bibinfo {author} {\bibfnamefont {B.}~\bibnamefont
  {Schwikowski}},\ }\bibfield  {title} {\enquote {\bibinfo {title} {Graph-based
  methods for analysing networks in cell biology},}\ }\href@noop {} {\bibfield
  {journal} {\bibinfo  {journal} {Briefings in bioinformatics}\ }\textbf
  {\bibinfo {volume} {7}},\ \bibinfo {pages} {243--255} (\bibinfo {year}
  {2006})}\BibitemShut {NoStop}%
\bibitem [{\citenamefont {Rahman}, \citenamefont {Bhuiyan},\ and\ \citenamefont
  {Hasan}(2012)}]{Graft}%
  \BibitemOpen
  \bibfield  {author} {\bibinfo {author} {\bibfnamefont {M.}~\bibnamefont
  {Rahman}}, \bibinfo {author} {\bibfnamefont {M.}~\bibnamefont {Bhuiyan}}, \
  and\ \bibinfo {author} {\bibfnamefont {M.~A.}\ \bibnamefont {Hasan}},\
  }\bibfield  {title} {\enquote {\bibinfo {title} {Graft: an approximate
  graphlet counting algorithm for large graph analysis},}\ }in\ \href@noop {}
  {\emph {\bibinfo {booktitle} {Proceedings of the 21st ACM international
  conference on Information and knowledge management}}}\ (\bibinfo
  {organization} {ACM},\ \bibinfo {year} {2012})\ pp.\ \bibinfo {pages}
  {1467--1471}\BibitemShut {NoStop}%
\bibitem [{Note1()}]{Note1}%
  \BibitemOpen
  \bibinfo {note} {Http://dblp.uni-trier.de/xml}\BibitemShut {NoStop}%
\bibitem [{Note2()}]{Note2}%
  \BibitemOpen
  \bibinfo {note} {Http://dblp.uni-trier.de/xml}\BibitemShut {NoStop}%
\bibitem [{Note3()}]{Note3}%
  \BibitemOpen
  \bibinfo {note} {Http://snap.stanford.edu}\BibitemShut {NoStop}%
\bibitem [{Note4()}]{Note4}%
  \BibitemOpen
  \bibinfo {note} {Http://snap.stanford.edu}\BibitemShut {NoStop}%
\bibitem [{Note5()}]{Note5}%
  \BibitemOpen
  \bibinfo {note} {Http://snap.stanford.edu}\BibitemShut {NoStop}%
\bibitem [{Note6()}]{Note6}%
  \BibitemOpen
  \bibinfo {note} {Http://konect.uni-koblenz.de}\BibitemShut {NoStop}%
\bibitem [{Note7()}]{Note7}%
  \BibitemOpen
  \bibinfo {note} {Http://deim.urv.cat/\textasciitilde aarenas}\BibitemShut
  {NoStop}%
\end{thebibliography}%

\end{document}